%% file: main.tex
\documentclass[%
superscriptaddress,
amsmath,amssymb,
reprint,
nofootinbib,
floatfix,
]{revtex4-2}
\usepackage{graphicx}
\usepackage{caption}
\usepackage{subcaption}
\usepackage{dcolumn}
\usepackage{placeins}

\usepackage[mathlines]{lineno}
\modulolinenumbers[5]
\usepackage{hyperref}
\hypersetup{
  colorlinks=true,
  citecolor=blue,
  linkcolor=blue,
  urlcolor=blue
}

\begin{document}

\title{Uncertainty Aware Learning for High Energy Physics}
\author{Aishik Ghosh}
\affiliation{Department of Physics and Astronomy, University of California, Irvine, CA 92697, USA}
 \affiliation{Physics Division, Lawrence Berkeley National Laboratory, Berkeley, CA 94720, USA}
\author{Benjamin Nachman}
 \affiliation{Physics Division, Lawrence Berkeley National Laboratory, Berkeley, CA 94720, USA}
 \affiliation{Berkeley Institute for Data Science, University of California, Berkeley, CA 94720, USA}
\author{Daniel Whiteson }
\affiliation{Department of Physics and Astronomy, University of California, Irvine, CA 92697, USA}

\begin{abstract}
Machine learning techniques are becoming an integral component of data analysis in High Energy Physics (HEP).  These tools provide a significant improvement in sensitivity over traditional analyses by exploiting subtle patterns in high-dimensional feature spaces. These subtle patterns may not be well-modeled by the simulations used for training machine learning methods, resulting in an enhanced sensitivity to systematic uncertainties.  

Contrary to the traditional wisdom of constructing an analysis strategy that is invariant to systematic uncertainties, we study the use of a classifier that is fully aware of uncertainties and their corresponding nuisance parameters.  We show that this dependence can actually enhance the sensitivity to parameters of interest.  Studies are performed using a synthetic Gaussian dataset as well as a more realistic HEP dataset based on Higgs boson decays to tau leptons.  For both cases, we show that the uncertainty aware approach can achieve a better sensitivity than  alternative machine learning strategies.

\end{abstract}

\maketitle

{\small
\tableofcontents
}

\section{Introduction}
\label{sec:Intro}

The usefulness of physical measurements is tied to the magnitude and reliability of their estimated uncertainties.  Whether one is measuring the mass of the Higgs boson~\cite{Aad:2015zhl} or the value of the Hubble constant~\cite{doi:10.1146/annurev-astro-082708-101829}, it is the size of the uncertainty that communicates the quality of the information and allows measurements to be contrasted or accurately combined.  While statistical uncertainties can be reduced with the collection of additional data, the most troublesome type of uncertainty is systematic uncertainty.  These uncertainties can arise from many sources and are often modeled as the dependence of a parameter of interest on other degrees of freedom known as \textit{nuisance parameters}. 

The challenging task of quantifying the systematic uncertainty on a measured physical parameter has become even more important due to the growing use of complex statistical procedures based on modern machine learning~\cite{Larkoski:2017jix,Guest:2018yhq,Albertsson:2018maf,Radovic:2018dip,Carleo:2019ptp,Bourilkov:2019yoi,Schwartz:2021ftp}. While more powerful techniques can extract more information from higher-dimensional datasets, they may also introduce or enhance the dependence of those measurements on nuisance parameters.  This is because, despite their importance, systematic uncertainties are often not part of the learning procedure for a typical machine learning model.  Instead, models are typically trained on synthetic datasets generated with assumed values of the nuisance parameters; the impact of uncertainties is typically quantified post-hoc by varying those nuisance parameters.   In cases where systematic uncertainties are dominant, such machine learning models may improve the statistical uncertainty, but increase the systematic uncertainty by exacerbating the dependence on nuisance parameters.  To minimize the total uncertainty,  it is essential to incorporate uncertainties directly into the learning procedure.

To date, several approaches have been considered. Data augmentation trains a model on a concoction of synthetic data with different values of the nuisance parameters. This exposes the classifiers to the range of possible nuisance parameter values, so that at inference time, the result may be less sensitive to their precise value. The disadvantage of augmentation is that the model is unaware of the values of the nuisance parameters, and so learns the average, rather than the optimal response.   Another possibility is to train a model to explicitly be insensitive to nuisance parameters or other parameters~\cite{Blance:2019ibf,Englert:2018cfo,Louppe:2016ylz,Dolen:2016kst,Moult:2017okx,Stevens:2013dya,Shimmin:2017mfk,Bradshaw:2019ipy,ATL-PHYS-PUB-2018-014,DiscoFever,Wunsch:2019qbo,Rogozhnikov:2014zea,10.1088/2632-2153/ab9023,clavijo2020adversarial,Kasieczka:2020pil,Kitouni:2020xgb,Estrade:2019gzk}. One implementation of this approach involves training two machine learning models at the same time: one that achieves the primary learning task and a second model that tries to learn information about the nuisance parameter/feature from the output of the classifier.  When this second model (\textit{adversary}) is unable to perform its task, then the primary classifier is insensitive, as desired.  The data augmentation and adversarial learning approaches will serve as important baselines in this paper.

Rendering a classifier insensitive to nuisance parameters may increase analysis precision and decrease analysis complexity, but it may also have undesirable consequences.  First, systematic uncertainties often involve guesswork; in many cases, the corresponding nuisance parameters do not have a strict statistical origin.  Reducing the sensitivity to a particular nuisance parameter may not eliminate the underlying uncertainty; instead, it may eliminate the only existing handle to probe the source of uncertainty.  For example, a measurement which used a model constructed to be insensitive to differences between two example parton shower algorithms, such as \textsc{Pythia}~\cite{Sjostrand:2007gs} and \textsc{Herwig}~\cite{Bellm:2015jjp}, may still have significant systematic uncertainty due to our lack of a complete understanding of fragmentation.  Second, some values of the nuisance parameter achieve a better sensitivity to the parameters of interest than others and this could be exploited to improve the performance of a classifier. A classifier that is insensitive to a nuisance parameter will not be able to exploit features that are highly sensitive to that parameter. The first example in this paper demonstrates an extreme case where a classifier insensitive to a nuisance parameter will result in no separation power.

We advocate for the opposite of decorrelation.  Classifiers are constructed to be explicitly dependent on nuisance parameters as if they are parameters of interest.  As nuisance parameters are profiled, the classifier will change and the best classifier will be used for each value of the nuisance parameter. Parameterized classifiers have been studied in the context of parameters or features of interest~\cite{Cranmer:2015bka,Baldi:2016fzo}, and full dependence on nuisance parameters for inference has been advocated in Ref.~\cite{Brehmer:2019xox,Brehmer:2018hga,Brehmer:2018kdj,Brehmer:2018eca,Nachman:2019dol}. 

In this paper, we provide specific examples of profiled classifiers and show explicitly that they can enhance analysis sensitivity over strategies that render networks insensitive to nuisance parameters. We focus on only the construction of classifiers as useful statistics for downstream analysis and not on full likelihood (ratio) estimation.  In this way, our uncertainty-aware classifier approach is a straightforward extension of existing analyses performed at the Large Hadron Collider (LHC) and therefore may result in immediate improvements in sensitivity. In addition, this prescription allows for easy post-hoc histogram-based diagnostics. These may include quantification of the impact of additional sources of systematic uncertainties that are not used for training, and checks for whether the measurement over-constrains the nuisance parameter.

While we focus on the profiling aspect of uncertainty awareness, there is a complementary line of research on the use of uncertainty-aware loss functions~\cite{Wunsch:2020iuh,Elwood:2020pik,Xia:2018kgd,deCastro:2018mgh,Charnock_2018,Alsing:2019dvb,lukas_heinrich_2020_3697981}.  These approaches construct classifiers that are optimized using the final test statistic, including uncertainties.  We leave the combination of profile-aware training and uncertainty-aware loss to future work.  There have also been recent proposals to use Bayesian neural networks for estimating uncertainties~\cite{Kasieczka:2020vlh,Bollweg:2019skg,Araz:2021wqm,Bellagente:2021yyh}.  Additional information about the interplay between uncertainties and machine learning can be found in recent reviews~\cite{Nachman:2019dol,1807719}.

This paper is organized as follows.  The uncertainty-aware methods are introduced in Sec.~\ref{sec:methods}.  Evaluation criteria for assessing the performance of the methods is described in Sec.~\ref{sec:approach}.  To build intuition, a Gaussian example is presented in Sec.~\ref{sec:Gaussian}.  A physics example based on Higgs boson decays to $\tau$ leptons using a benchmark dataset is provided in Sec.~\ref{sec:Physics}.  The paper ends with conclusions and outlook in Sec.~\ref{sec:conclusions}.

\section{Uncertainty-Aware Methods}
\label{sec:methods}

This section describes the four methods of training classifiers studied in this paper. All neural networks were trained using \textsc{Keras 2.2.4}~\cite{chollet2015keras} with a \textsc{TensorFlow 1.12.0}~\cite{tensorflow2015-whitepaper} backend on a single \textsc{Nvidia GeForce GTX} GPU.

\subsection{Notation}

Features used for classification are denoted by $X\in\mathbb{R}^n$, where lower case $x$ is a realization of the random variable $X$.  A dataset of many examples will be written $\{x_i\}$.  The nominal value of the nuisance parameter $z\in\mathbb{R}^m$ is $z_0$ and the true value is $z_\textrm{T}$.  In some cases, we will promote the nuisance parameters to random variables, in which case they will be represented by the capital letter $Z$.

\subsection{Baseline Classifier}
The baseline method is a classifier trained to distinguish signal and background  using data simulated at the nominal value of the nuisance parameter, as is done routinely in LHC analyses. A network trained optimally to minimise a Binary Cross-Entropy (BCE) loss learns to output a score (see e.g.~\cite{hastie01statisticallearning}),
\begin{equation}
    s(x) = \frac{p(x|z=z_{0},S)}{p(x| z=z_{0},S) + p(x| z=z_{0},B)}\,,
\label{eq:baseline_score}
\end{equation}
where 

$p(\cdot)$ denotes a probability density, $S$ represents the signal class and $B$ represents the background class.
The score of the network is used as an observable with high sensitivity to the parameter of interest for the final measurement.

\subsection{Data Augmentation}
An alternative method is to augment the training data to include signal and background samples with several values of the nuisance parameters. A network trained optimally to minimise a BCE loss learns the score,

\begin{equation}
    s(x) = \frac{\langle p(x|Z,S)\rangle_{p_Z}}{\langle p(x| Z,S)\rangle_{p_Z} + \langle p(x| Z,B)\rangle_{p_Z}}\,,
\end{equation}
where $p_Z$ is the probability density over the nuisance parameter $Z$, treated as a random variable with some probability density chosen by the experimenter.  Typically, $Z$ is discrete and has a nonzero probability mass at only a few values.  The score $s(x)$ is then treated in the same way as in the baseline case (Eq.~\ref{eq:baseline_score}).

\subsection{Adversarial Training}
An orthogonal strategy is to train a classifier with the explicit objective of being insensitive to the effects of the nuisance parameter. Our implementation follows the adversarial training prescription of Ref.~\cite{Louppe:2016ylz}. However, to improve the training stability and speed, the classifier and adversary are concatenated together through a gradient reversal layer~\cite{2014arXiv1409.7495G} and trained simultaneously. The classifier is trained with the objective to minimize the classification loss and maximise the adversarial loss and the second loss has a relative weight of 
$\lambda$, a tunable hyper-parameter. 

While training for exact invariance in this adversarial setup can be tricky~\cite{Estrade:DLPS2017}, maximizing overall sensitivity requires a compromise between the level of invariance to nuisance parameters and the classification power. The Gaussian case described in Sec.~\ref{sec:Gaussian} is an extreme example where exact invariance to the nuisance parameter requires zero discriminating power for the classifier.

In the end, the score of the classifier on observed data is used as an observable in the final measurement, in the same way as for the baseline classifier.

\subsection{Uncertainty-Aware Classifier}
\begin{figure}[ht]
  \centering
    \includegraphics[width=0.35\textwidth]{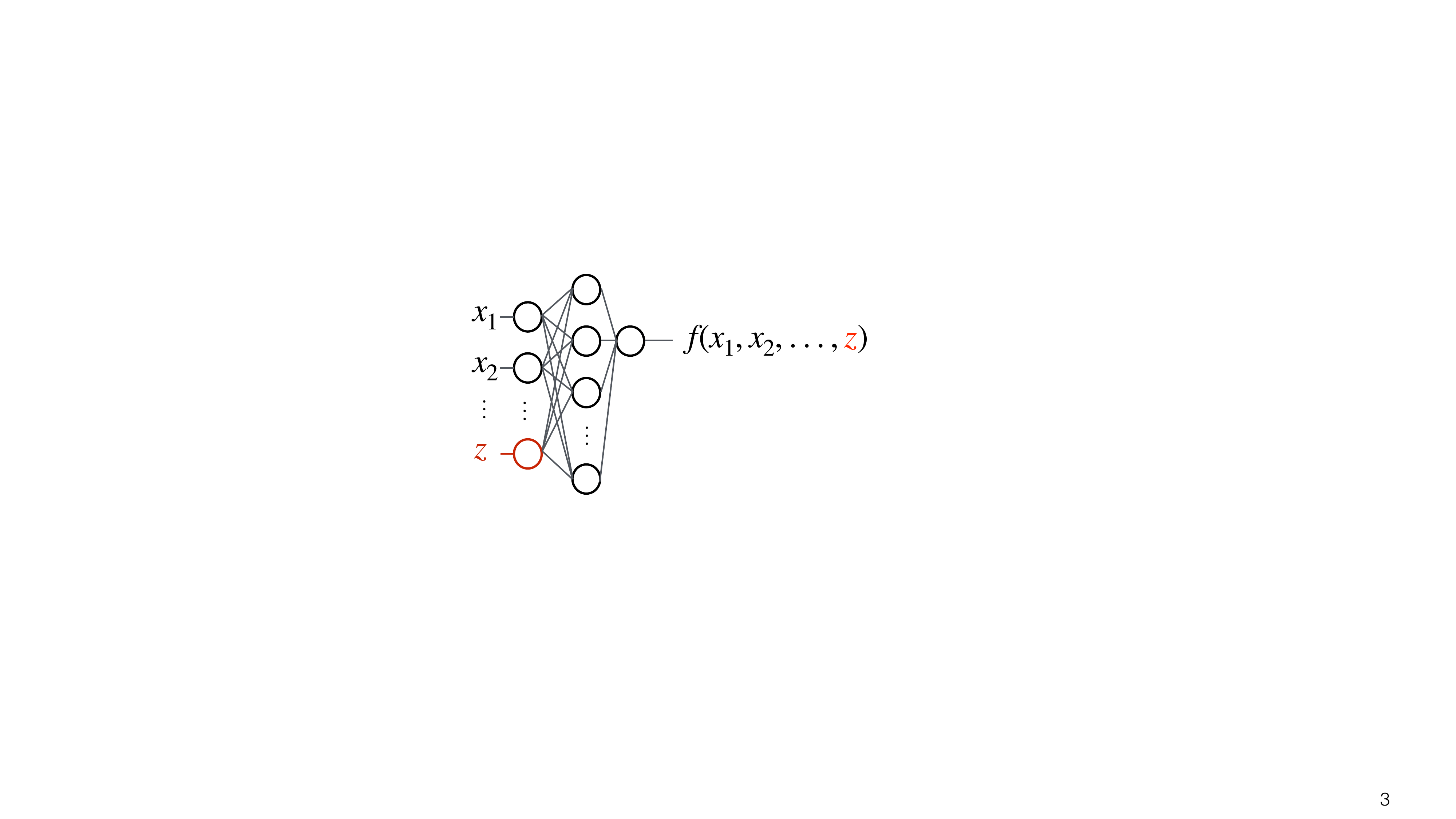}
  \caption{The architecture of an uncertainty-aware network, in which the nuisance parameter $z$ is treated as a feature alongside the observed data $x$, learning a decision function which varies with the nuisance parameter.}
  \label{fig:NN_arch}
\end{figure}

The concept explored in this paper is to parameterize the network in the nuisance parameters; see Fig.~\ref{fig:NN_arch}. Specifically, the network is trained with the true value of the nuisance parameter $z$ as an input to the network in additional to the observables $x$. A network trained optimally to minimise a BCE loss learns the score,
\begin{equation}
    s(x,z) = \frac{p(x|Z=z,S)}{p(x|Z=z,S) + p(x|Z=z,B)}.
\label{eq:aware_score}
\end{equation}

The score of this classifier is not used as a single observable for the final fit as in the previous methods. At evaluation time, while the $x$ values remain fixed as inputs to the network, the unknown $z$ is left as a parameter, allowing for later profiling over 
the nuisance parameters in the final measurement.

Importantly, note that Eq.~\ref{eq:aware_score} depends on $z$.  This means that the calculation of analysis observable(s) depends on $z$ and change as the nuisance parameter is varied, during the evaluation of uncertainties and/or during nuisance parameter profiling. This is in contrast to the standard search paradigm in which the calculation of the analysis observables are fixed and the sensitivity to $z$ is evaluated post-hoc.   Allowing the calculation of the analysis observables to depend explicitly on the value of $z$ is not the traditional approach, but it does not require that the experimenter have any special knowledge of $z$. Formation of a confidence interval in the space of model parameters (either parameters of interest or nuisance parameters) naturally requires calculating the likelihood ratio of the model as those parameters vary, relative to the best-fit parameters. It is natural for the calculation of the analysis observable, a proxy for the likelihood ratio, to vary with those parameters. One can later profile over the nuisance parameters to capture the impact of our lack of knowledge of its true value. The traditional approach of fixing the analysis observable calculation can be thought of as an ad-hoc approximation of the full method.

\section{Evaluation Methodology}
\label{sec:approach}

To evaluate the power of each approach above, we apply them to a common use case, fitting a signal hypothesis in the presence of background, where both signal and background depend on nuisance parameters.  Relevant to many measurements of Standard Model (SM) processes as well as searches for physics beyond the SM, the parameter of interest is the signal strength $\mu$, the cross section of the signal relative to the reference value. In the Gaussian example below, we use low-dimensional datasets for simpler visualization, but the results generalize. Similarly, for ease of calculations we perform a binned likelihood fit, although the unbinned nature of neural networks should allow application to unbinned cases; we leave that investigation to future work.

For each of the strategies described, template histograms of the classifier score are constructed from  simulated signal and background events for several values of the nuisance parameter $z$. These templates are the basis of the binned likelihood calculation $\mathcal{L}(\mu,z | \{x_i\})$ over the parameters ${\mu,z}$, where $\{x_i\}$ is the full observed dataset. The likelihood is a product of a Poisson term for each histogram bin  and a Gaussian constraint on the nuisance parameter. The Gaussian constraint can readily be replaced with any other prior or a Poisson term from an auxiliary measurement if $z$ is directly constrained with control region data (demonstrated in Appendix~\ref{sec:app:AuxMeasure}). If no additional prior or constraint on the nuisance parameter is used then only information from the primary measurement constrains $z$.  The Negative Log-Likelihood (NLL) is (up to an irrelevant constant),
\begin{align}\nonumber
    -&\log{\mathcal{L}(\mu,z | \{x_i\})}\\ \label{eq:2D_NLL}\nonumber &=-\sum^{n_\text{bins}}_{j=1}\bigg[ N_j\cdot \log{(\mu s_j + b_j)} - \mu s_j -b_j - \log(\Gamma(N_i)) \bigg] \\
    &\qquad+  \left(\frac{z - z_{0}}{\sqrt{2}\sigma_z} \right)^2,
\end{align}
where $s_j$, $b_j$ are the expected number of signal and background events in bin $j$, respectively, and $N_j$ is the number of events observed in data for that bin.  The $\Gamma$ function is the generalized factorial function which can handle decimal values in the simulated test dataset. Although usually irrelevant, the $\log(\Gamma(N_i))$ term is not a constant while using an uncertainty-aware network and cannot be ignored. For this approach, the decision function changes with $z$ and therefore the bin counts in simulation and observed data also change with $z$. 

In practice, samples at various values of $z$ can often be produced cheaply from a single simulated MC sample by shifting the value of $z$ and recomputing all the relevant physics variables, and this approach will be used for the studies in Sec.~\ref{sec:Physics}. Care must be taken to apply any kinematic selection on these variables only after the shift. In these studies, the templates and the `observed dataset' are built using the same test dataset because the dataset used in Sec.~\ref{sec:Physics} is not large enough to split into three representative datasets.

The fitted value of $\mu$ is obtained by minimizing Eq.~\ref{eq:2D_NLL}.   Uncertainties are accounted for by studying the dependence of the likelihood near the fitted value $\hat{\mu}$ while optimizing over $z$. The power of each approach is determined by their relative uncertainties in $\mu$. As a diagnostic, the parameter of interest may be profiled over instead to check if the measurement over-constrains the nuisance parameter.


\section{Gaussian Example}
\label{sec:Gaussian}

\begin{figure}[b!p]
  \centering
    \includegraphics[width=0.45\textwidth]{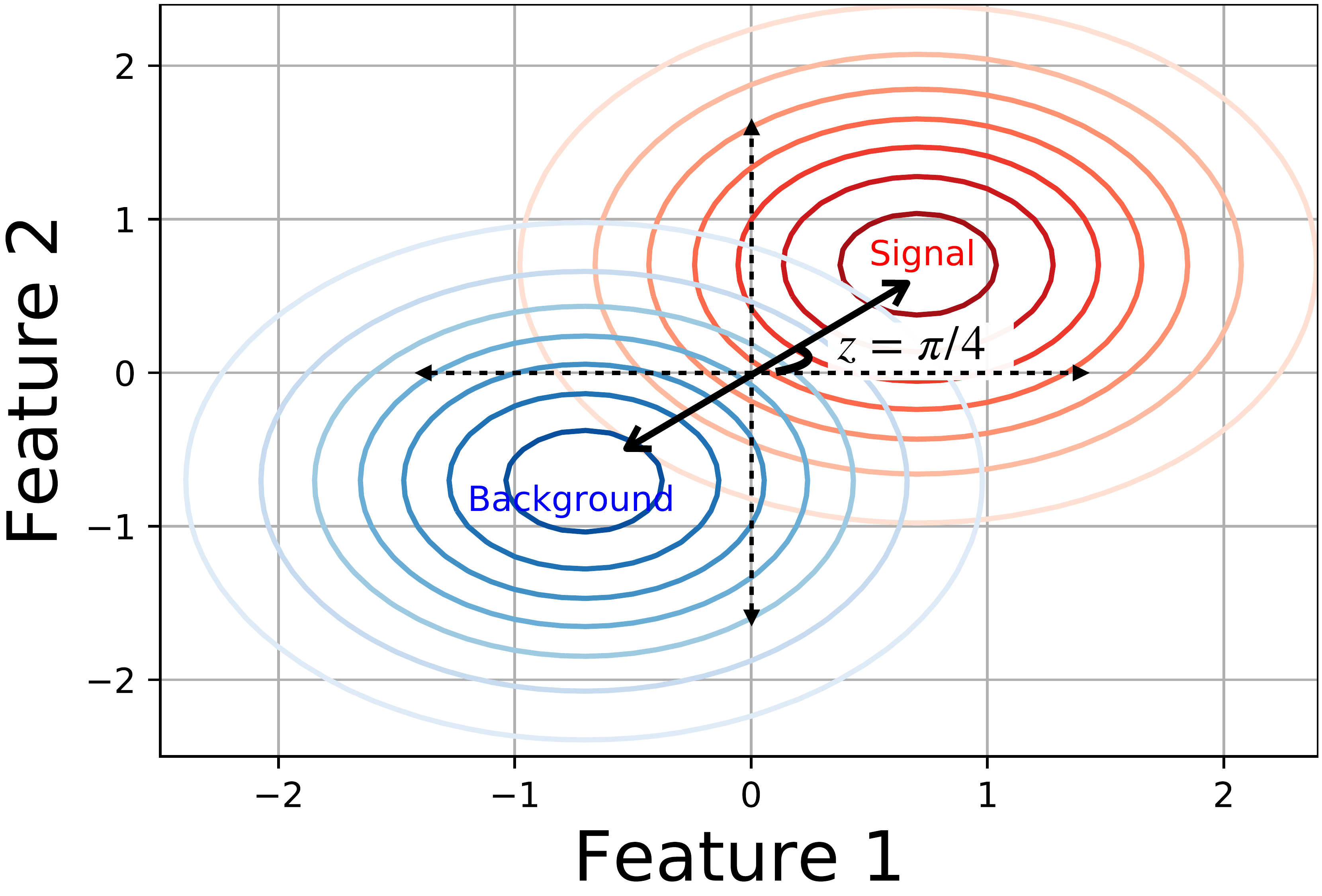}
  \caption{Contour of probability densities for signal and background hypotheses in the two-dimensional feature space for the simple Gaussian demonstration case, with the nuisance parameter fixed to  $z=\frac{\pi}{4}$. }
  \label{fig:ToyData}
\end{figure}

To illustrate the different approaches in a simple setting with complete analytic control, we begin with a Gaussian example with a two-dimensional feature space and a single nuisance parameter.  Signal events are drawn from Gaussian distributions in the two features, with means at $\cos{(z)}$ and $\sin{(z)}$, respectively; the width of each is set to $0.7$. Background events are generated in same fashion, but with means for the two features at $-\cos{(z)}$ and $-\sin{(z)}$ respectively. An example of the signal and background distributions for $z=\frac{\pi}{4}$ is shown in Fig.~\ref{fig:ToyData}.

A set of $4.2\times 10^7$ events are generated at 21 values of $z$ equally spaced between $0$ and $\pi/2$.  The dataset is split into training and test sets with a ratio of 3:1. All signal events in the test set have a weight of $10^{-3}$ and all background events have a weight of $10^{-1}$ to mimic a rare signal typical of LHC analyses. Ten bins are used to construct the template and observed histograms. The parameter of interest is the signal strength $\mu$ with a true value of $1$. 

\subsection{Models}
In a simple case where the signal and background probabilities are well known, it is possible to derive  the classifier analytically for the baseline  and uncertainty-aware approaches. The results below use the analytical expressions, but as a cross check, neural networks were also trained for the same objective and produced nearly identical results.
\subsubsection{Baseline and Uncertainty-Aware Classifiers}
The baseline classifier computes the score
\begin{equation}
    s(x)=\frac{p(x|z=\frac{\pi}{4},S)}{p(x| z=\frac{\pi}{4},S) + p(x|z=\frac{\pi}{4},B)}\,,
\end{equation}
using the the probability density functions for the Gaussian distributions used to generate the two features for signal and background at an assumed fixed value of $z=\frac{\pi}{4}$. 

The uncertainty-aware classifier, on the other hand, does not make  assumptions about the value of the nuisance parameter and instead calculates a score as a function of the nuisance parameter
 
\begin{equation}
s(x,z)=\frac{p(x|Z=z,S)}{p(x|Z=z,S) + p(x|Z=z,B)}.
\end{equation}

\begin{figure}[ht]
  \centering
  \begin{subfigure}[b]{0.23\textwidth}
    \centering
    \includegraphics[width=\textwidth]{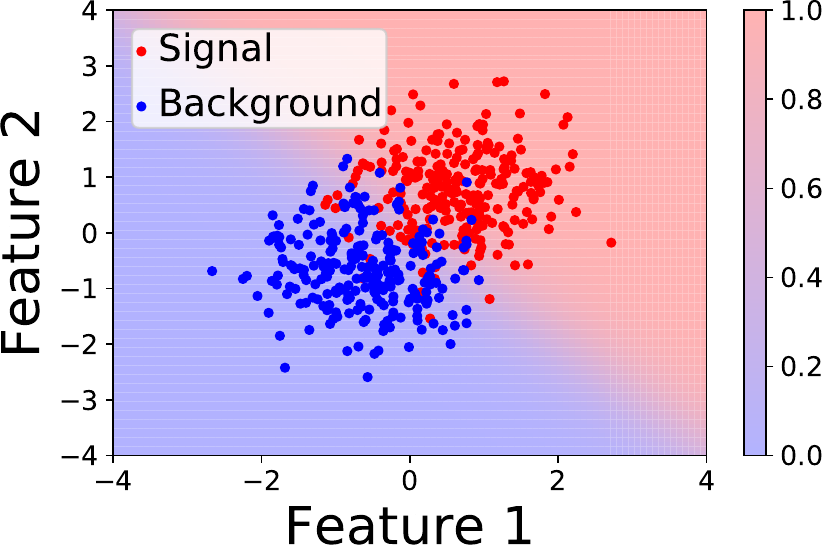}
    \caption{Baseline, assuming $z=\frac{\pi}{4}$, on data where $z=\frac{\pi}{4}$}
    \label{fig:df:baseline_nom}
  \end{subfigure}
  \hfill
  \begin{subfigure}[b]{0.23\textwidth}
    \centering
\includegraphics[width=\textwidth]{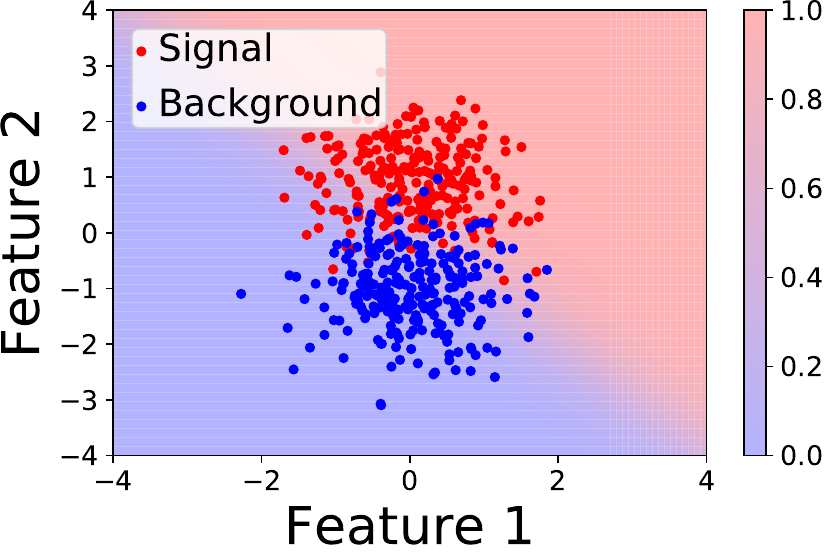}
        \caption{Baseline, assuming $z=\frac{\pi}{4}$, on data where $z=\frac{\pi}{2}$}
    \label{fig:df:baseline_systUp}
  \end{subfigure}
  \hfill
  \begin{subfigure}[b]{0.23\textwidth}
    \centering
    \includegraphics[width=\textwidth]{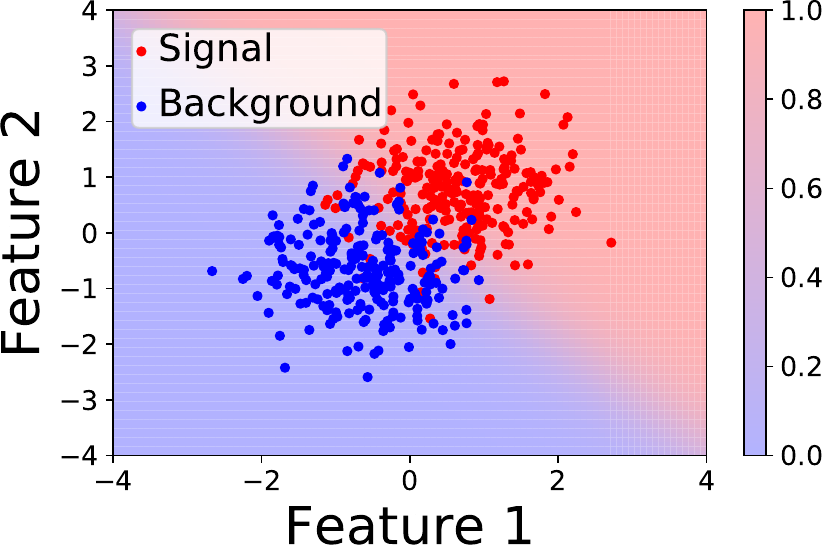}
     \caption{Uncertainty-aware, evaluated at $z=\frac{\pi}{4}$, on data where  $z=\frac{\pi}{4}$}
    \label{fig:df:aware_nom}
  \end{subfigure}
  \hfill
  \begin{subfigure}[b]{0.23\textwidth}
    \centering
\includegraphics[width=\textwidth]{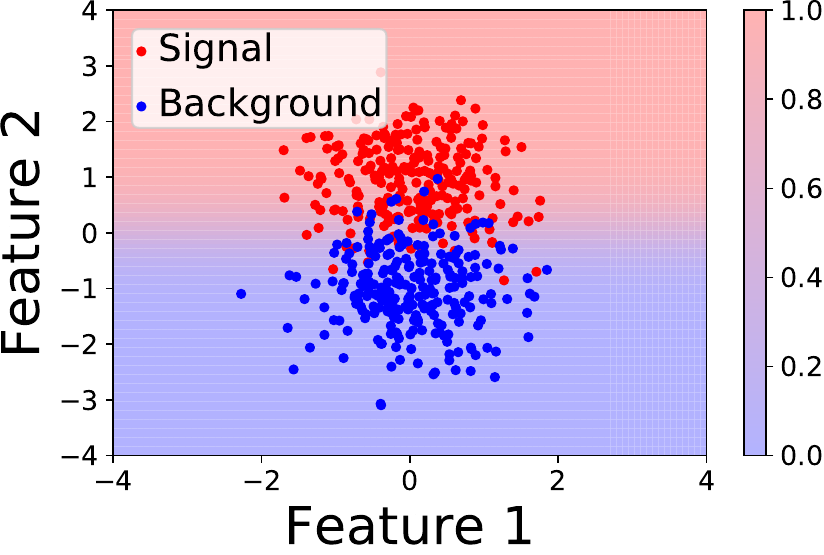}
\caption{Uncertainty-aware, evaluated at $z=\frac{\pi}{2}$, on data generated with $z=\frac{\pi}{2}$}    \label{fig:df:aware_SystUp}
  \end{subfigure}
  \caption{ Classifier score for the baseline and systematic-aware classifiers, see text for definitions.  Shown are examples where the baseline classifier's assumption that the nuisance parameter is $z=\frac{\pi}{4}$ matches or disagrees with the generated data (points). Also shown are score functions for the uncertainty-aware classifier on the same datasets, evaluated at the correct value of $z$ for each dataset. }
  \label{fig:decision_fuctions}
\end{figure}

\begin{figure}[b!p]
  \centering
  \begin{subfigure}[b]{0.23\textwidth}
    \centering
    \includegraphics[width=\textwidth]{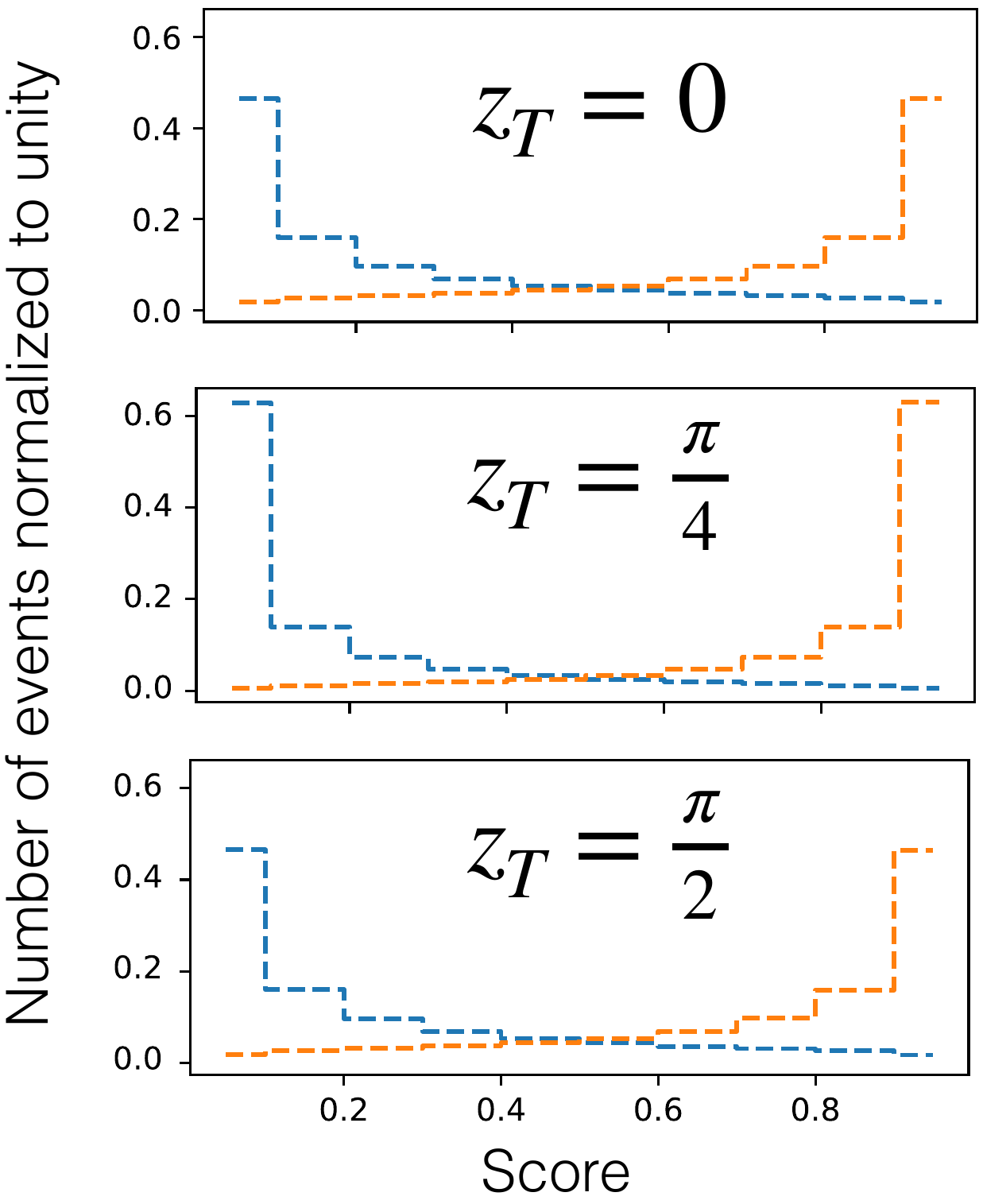}
    \caption{Baseline}
    \label{fig:templates:baseline}
  \end{subfigure}
  \hfill
  \begin{subfigure}[b]{0.23\textwidth}
    \centering
\includegraphics[width=\textwidth]{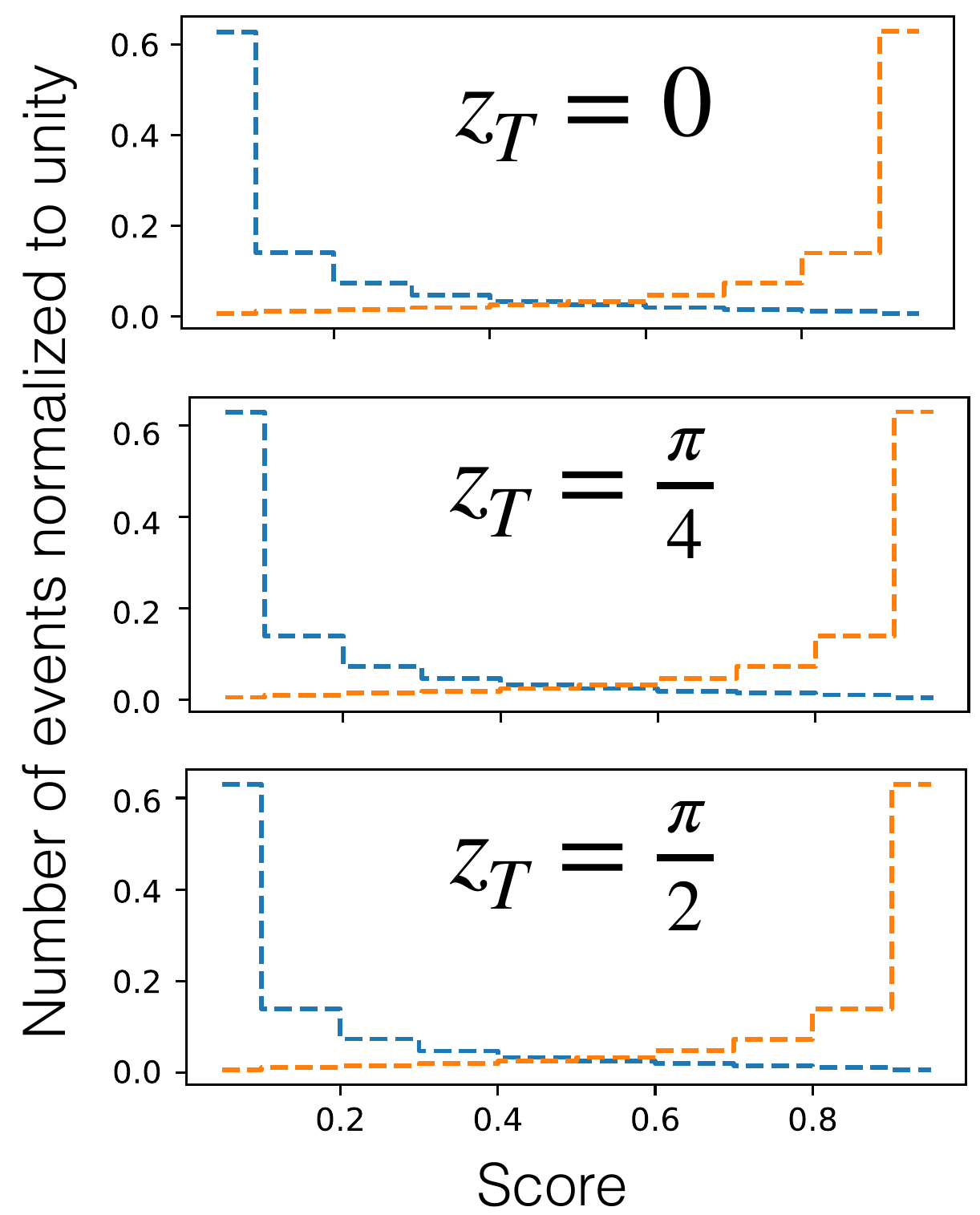}
    \caption{Uncertainty Aware}
    \label{fig:templates:aware}
  \end{subfigure}
  \caption{Template histograms of the classifier score for the baseline (left) and uncertainty-aware approaches (right) evaluated for data generated at various true values of $z$. The signal distribution is shown in orange and the background distribution in blue. The baseline classifier assumes $z=\frac{\pi}{4}$, and loses separation power for data generated with $z=\{0, \frac{\pi}{2}\}$, manifested by the lower heights of the signal and background histograms near $1$ and $0$, respectively. The uncertainty-aware classifier score is evaluated for the correct value of $z$, providing the optimal score in each case.  }
  \label{fig:templates}
\end{figure}

The score $s$, for the each of the two classifiers are shown in Fig.~\ref{fig:decision_fuctions} as a function of the input features, for datasets generated with $z=\frac{\pi}{4}$ or $z=\frac{\pi}{2}$. The uncertainty-aware classifier is parameterized as a function of $z$, and given the correct value of the nuisance parameter, it can provide the appropriate classifier. Examples of histogram templates of the classifier outputs are shown in Fig.~\ref{fig:templates}. The separation power of the baseline classifier is clearly reduced for cases where the data are generated with values of the nuisance parameter which do not match its assumed value of $z=\frac{\pi}{4}$. Using the Area Under the Receiver Operating Characteristic Curve as a metric to quantify separation power of a model, the separation power for the baseline classifier falls from $0.978$ for data generated at $z=\frac{\pi}{4}$ to $0.924$ for data generated at $z=\frac{\pi}{2}$, while it remains $0.978$ on both datasets for the uncertainty-aware classifier.

\subsubsection{Data Augmentation}
A Linear Discriminant Analysis (LDA) classifier from Scikit-Learn~\cite{scikit-learn} is trained on a  training dataset which includes samples with all 21 values\footnote{The data augmentation classifier was also trained on a dataset with a continuous distribution of $z$ sampled from the Gaussian prior of $z$ and found to provide near identical results.} of $z$. As a cross check, a neural network was trained on the same data and produced a nearly identical score function.

\subsubsection{Adversarial Training}
The adversarial architecture was trained using samples from all 21 values of $z$. The classifier and the adversarial network each consist of 10 hidden layers with 64 nodes and a rectified linear unit (ReLU) activation and a single node output layer with sigmoid and linear activations respectively. An L2 kernel regularizer~\cite{cortes2012l2} was applied to all but the first and final layer of each network. The two networks were attached with a gradient reversal layer which scales the gradient by $-0.2$ and trained with the RMSProp~\cite{Tieleman2012} optimizer and a batch size of 4096. BCE is used as the classification loss while Mean Squared Error (MSE) is used for the loss of the adversary. An adversarial loss weight of $\lambda=1$ was used. For this dataset, a classifier exactly invariant to $z$ would have zero separation power between signal and background. Therefore, a compromise between invariance and classification power was made in model selection, finding the largest value of $\lambda$ which did not deteriorate performance. Minimal hyper-parameter tuning was performed beyond tuning $\lambda$.

\subsection{Results}

\begin{figure}[b!p]
  \centering
  \begin{subfigure}[b]{0.23\textwidth}
    \centering
    \includegraphics[width=\textwidth]{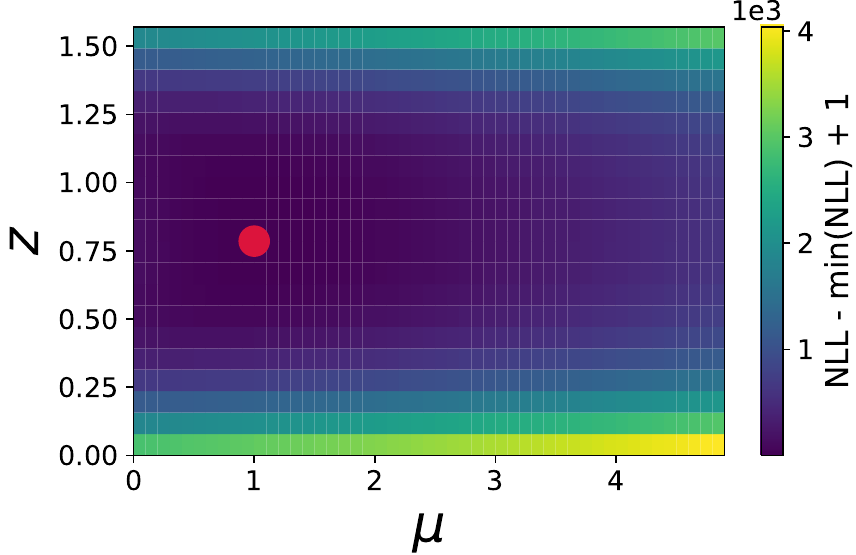}
    \caption{Baseline, assuming $z=\frac{\pi}{4}$, on data where $z=\frac{\pi}{4}$}
    \label{fig:2DNLLToy:baseline_nom}
  \end{subfigure}
  \hfill
  \begin{subfigure}[b]{0.23\textwidth}
    \centering
    \includegraphics[width=\textwidth]{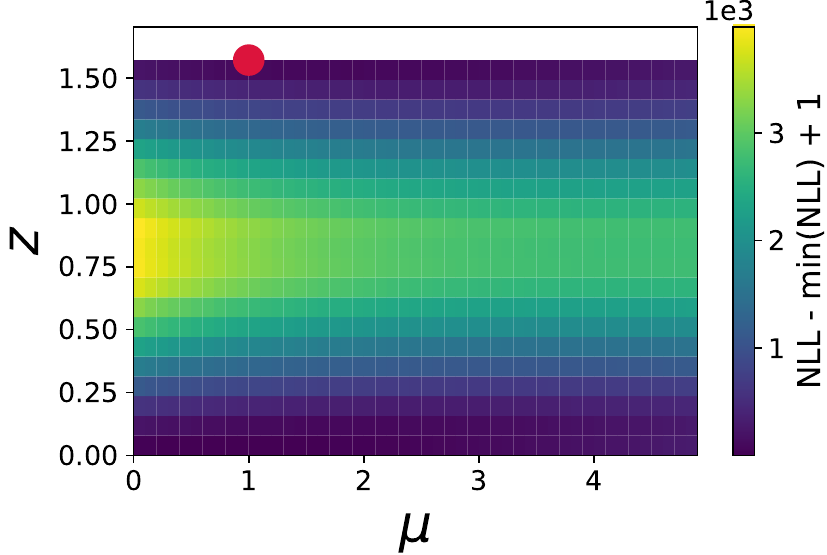}
    \caption{Baseline, assuming $z=\frac{\pi}{4}$, on data where $z=\frac{\pi}{2}$}
    \label{fig:2DNLLToy:baseline_up}
  \end{subfigure}
  \hfill
  \begin{subfigure}[b]{0.23\textwidth}
    \centering
\includegraphics[width=\textwidth]{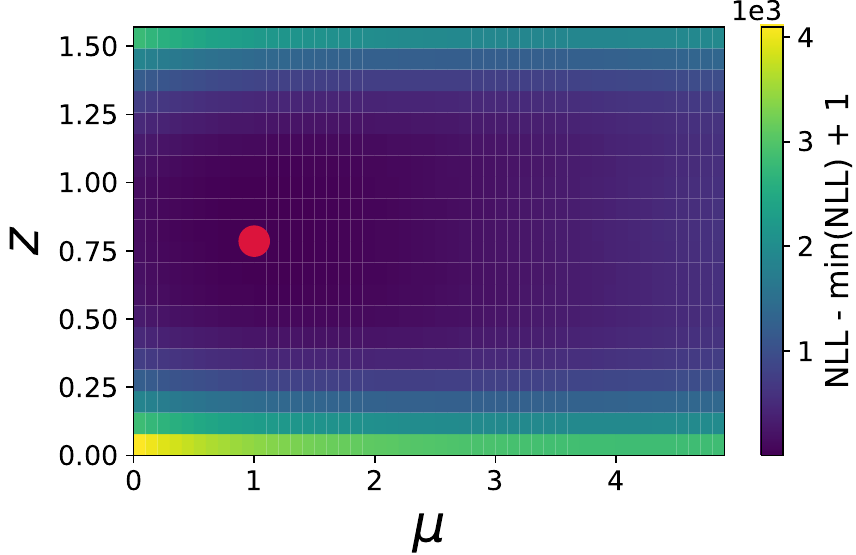}
     \caption{Uncertainty-aware, on data where  $z=\frac{\pi}{4}$}
    \label{fig:2DNLLToy:aware_nom}
  \end{subfigure}
  \hfill
  \begin{subfigure}[b]{0.23\textwidth}
    \centering
\includegraphics[width=\textwidth]{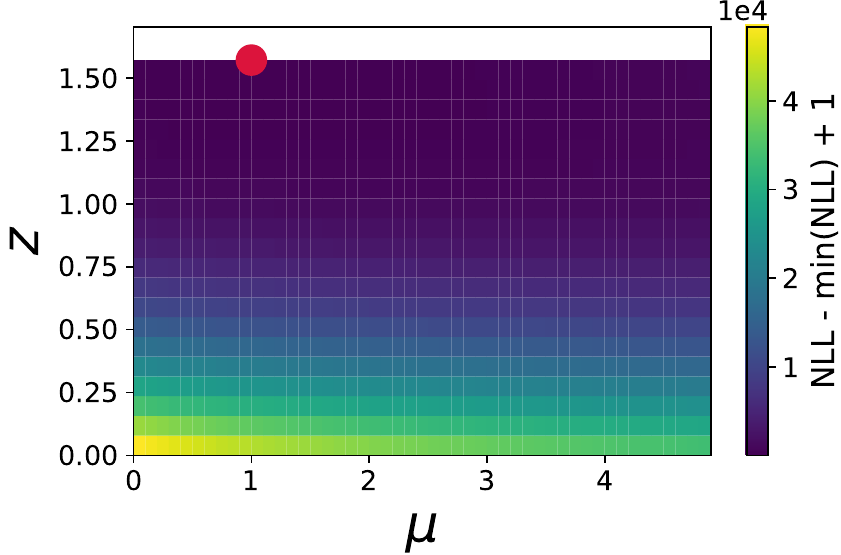}
     \caption{Uncertainty-aware, on data where  $z=\frac{\pi}{2}$}
    \label{fig:2DNLLToy:aware_up}
  \end{subfigure}
  \caption{ The negative log-likelihood (Eq.~\ref{eq:2D_NLL}) as a function of the parameter of interest $\mu$ and the nuisance parameter $z$ for two example datasets, using templates from the baseline (top) and uncertainty-aware classifier (bottom).  In the left column, the data are generated with $z=\frac{\pi}{4}$, which matches the assumption made by the baseline classifier.  In the right column, the data are generated with $=\frac{\pi}{2}$.  The red dot indicates the maximum likelihood estimate which coincides with the true value of $\mu, z$ in each case. Note the different z-axis scales for the two classifiers in the bottom row. }
  \label{fig:2DNLLToy}
\end{figure}

The negative log-likelihood (Eq.~\ref{eq:2D_NLL}) is calculated as a function of the parameter of interest $\mu$ and the nuisance parameter $z$.  Examples are shown in Fig.~\ref{fig:2DNLLToy} using templates from the baseline and uncertainty-aware classifiers. Due to its assumption that $z=\frac{\pi}{4}$ in the calculation of the classifier score, the likelihood from the baseline classifier can strongly exclude  $z=\frac{\pi}{4}$ when evaluated on a dataset generated with $z=\frac{\pi}{2}$, but finds $z=0, \frac{\pi}{2}$ equally likely. The uncertainty-aware classifier, on the other hand, is also able to exclude the low $z$ region.

 The measurement of the nuisance parameter is not the final objective and it can be profiled away. The most relevant metric for determining the relative power of the various approaches is the profile likelihood, $\max_z\mathcal{L}(\mu,z)$.
 
\begin{figure}[b!p]
  \centering
  \begin{subfigure}[b]{0.45\textwidth}
    \centering
    \includegraphics[width=\textwidth]{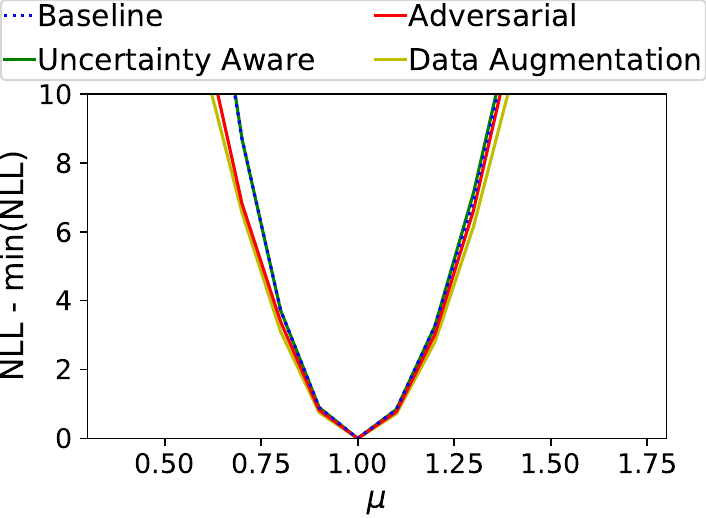}
    \caption{Data generated with $z=\frac{\pi}{4}$.}
    \label{fig:1DNLLToy:Nom}
  \end{subfigure}
  \hfill
  \begin{subfigure}[b]{0.45\textwidth}
    \centering
\includegraphics[width=\textwidth]{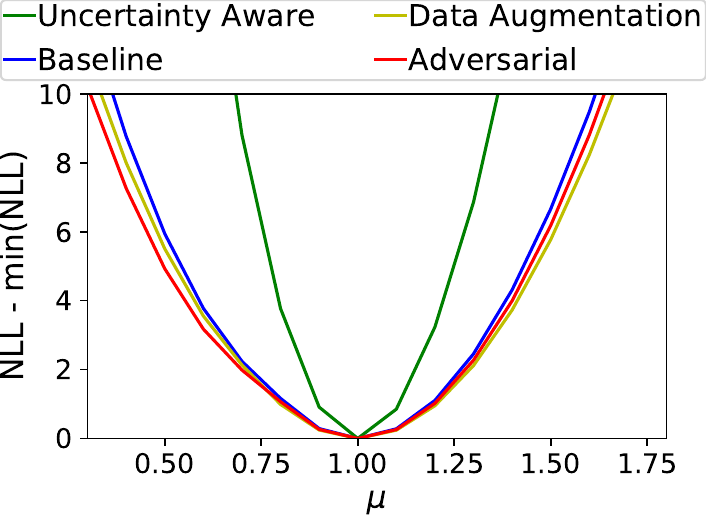}
    \caption{Data generated with $z=\frac{\pi}{2}$.}
    \label{fig:1DNLLToy:SystUp}
  \end{subfigure}
  \caption{ The profile likelihood $\max_z\mathcal{L}(\mu,z)$ as a function of the parameter of interest, $\mu$ for likelihoods calculated with templates built from the various classifiers. Narrower curves indicate more precise measurements having accounted for systematic and statistical uncertainties. The baseline classifier assumes $z=\frac{\pi}{4}$, and matches the performance of the uncertainty-aware classifier in data generated with $z=\frac{\pi}{4}$ (top).  In data generated with $z=\frac{\pi}{2}$, the power of all classifiers other than the uncertainty-aware classifier become significantly weaker.
    }
  \label{fig:1DNLLToy}
\end{figure}

The profile likelihood for each method is shown in Fig.~\ref{fig:1DNLLToy} for data generated with $z=\frac{\pi}{4}$ and $z=\frac{\pi}{2}$.  In the case of $z=\frac{\pi}{4}$, which matches the assumption of the baseline classifier, the uncertainty-aware and baseline classifiers both achieve ideal performance. The adversarial and data-augmentation approaches are somewhat weaker due to the inherent compromises of their methods.

When evaluated on data generated with $z=\frac{\pi}{2}$, in conflict with the assumption of the baseline classifier, the performance of all approaches other than the uncertainty-aware classifier deteriorate significantly. The data-augmented classifier has been trained on 21 values of $z$ in the first quadrant centred around the nominal value which makes it perform worse at extreme values of $z$. No setting of the adversarially-trained classifier was found to perform well for  datasets with both values of $z$.

\section{Realistic Example}
\label{sec:Physics}
A more realistic application of the uncertainty-aware classifier in the presence of nuisance parameters can be performed using the datasets~\cite{HiggsMLOpenData} produced for  the HiggsML Kaggle challenge~\cite{pmlr-v42-cowa14} by the ATLAS Collaboration. This dataset was originally simulated by the ATLAS collaboration to measure the decay of the Higgs boson to a pair of $\tau$~leptons~\cite{Aad:2015vsa}. This dataset was chosen for our study because it has been used as a benchmark for uncertainty aware learning in the past~\cite{Estrade:DLPS2017,Estrade_epjconf_2019}.

The signal process is the production of Higgs bosons through gluon-gluon fusion (ggF), vector boson fusion (VBF), and associated production with a vector boson (VH), which decay to pairs of $\tau$~leptons. The ggF and VBF production processes were simulated with \textsc{Powheg}~\cite{Nason:2004rx,Frixione:2007vw,Alioli:2010xd,Bagnaschi:2011tu} interfaced to \textsc{Pythia8}~\cite{Sjostrand:2007gs} while the VH production is simulated with \textsc{Pythia8}. Further details on corrections applied can be found in Sec~3. of Ref.~\cite{Aad:2015vsa}.
The detector response is simulated with GEANT4~\cite{Agostinelli:2002hh} and object reconstruction performed with the official ATLAS software~\cite{atlas_collaboration_2019_2641997}. The three largest backgrounds from $Z/\gamma^*\rightarrow \tau\tau$, $t\bar{t}$ and $W+\text{jets}$ are simulated with the same chain and mixed in proportions determined by their relative cross sections. Different aspects of the $Z/\gamma^*\rightarrow \tau\tau$ background are simulated with \textsc{Alpgen}, \textsc{Pythia8}, \textsc{Herwig}, and \textsc{Sherpa}~\cite{Gleisberg:2008ta}; the details can be found in Table~1 of Ref.~\cite{Aad:2015vsa}. The $t\bar{t}$ background is simulated with \textsc{Powheg} and \textsc{Pythia8} and the $W+\text{jets}$ background is simulated with \textsc{Alpgen}~\cite{Mangano:2002ea} and \textsc{Pythia8}.
Each event is characterized by 29 features\footnote{The DER\_mass\_MMC feature listed in Ref.~\cite{HiggsMLOpenData} was not included in the studies, following precedent set by Ref.~\cite{Estrade:DLPS2017}, because the Missing Mass Calculator~\cite{Elagin:2010aw} is slow to run and as an MCMC algorithm, introduces an additional source of stochasticity which makes comparisons difficult.}, including the lepton momenta and angles, the magnitude and direction of missing transverse momentum, the energy and angles of leading and sub-leading jets, and several other primary and derived variables. See Ref.~\cite{HiggsMLOpenData} for details.

The most important nuisance parameter is the unknown absolute energy scale of the hadronically decaying $\tau$~leptons. We follow prior studies~\cite{Estrade:DLPS2017,Estrade_epjconf_2019} and model this using a skewing function~\cite{victor_estrade_2018_1887847}
which is applied to the $\tau$~lepton $E_{\textrm{T}}$, for signal and background alike.  The minimum $E_\textrm{T}$ threshold of 22 GeV is applied after skewing.

\begin{figure}[b!p]
  \centering
  \begin{subfigure}[b]{0.4\textwidth}
    \centering
    \includegraphics[width=\textwidth]{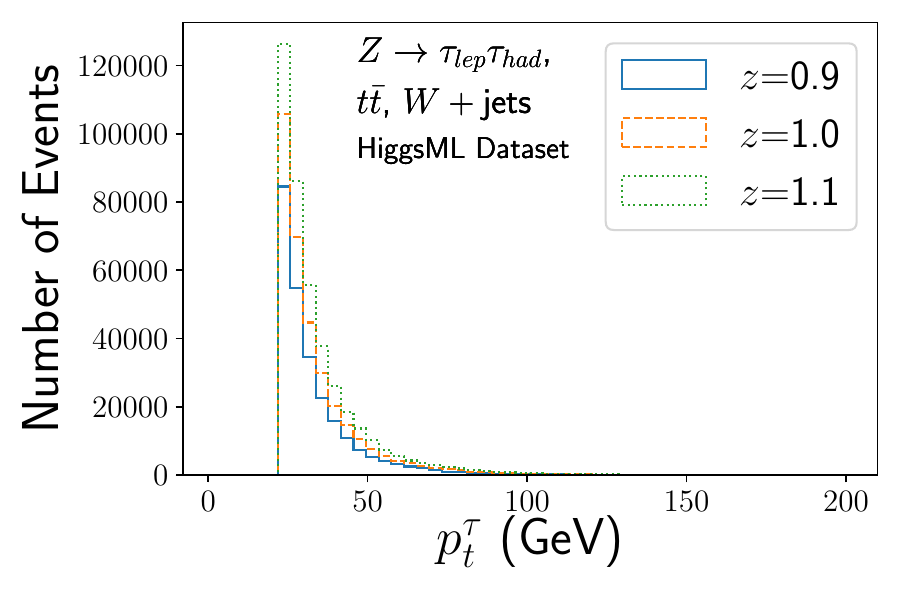}
    \caption{$p_t^\tau\text{ (GeV)}$}
    \label{fig:HiggsMLVariables:TauPt_0}
    \end{subfigure}
  \hfill
  \begin{subfigure}[b]{0.4\textwidth}
    \centering
\includegraphics[width=\textwidth]{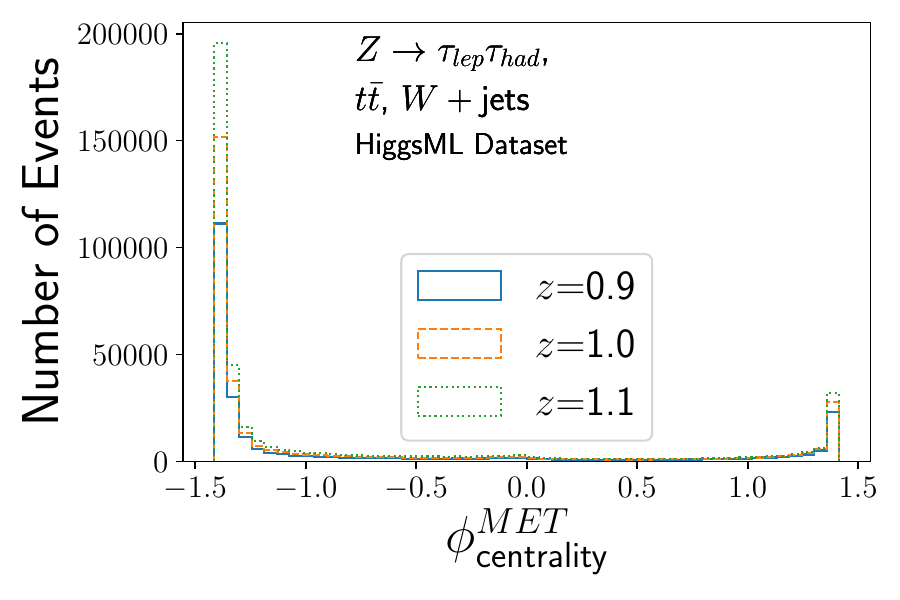}
    \caption{$\phi_{\text{centrality}}^{MET}$}
    \label{fig:HiggsMLVariables:phiCentralityMET_0}
  \end{subfigure}
    \hfill
  \begin{subfigure}[b]{0.4\textwidth}
    \centering
\includegraphics[width=\textwidth]{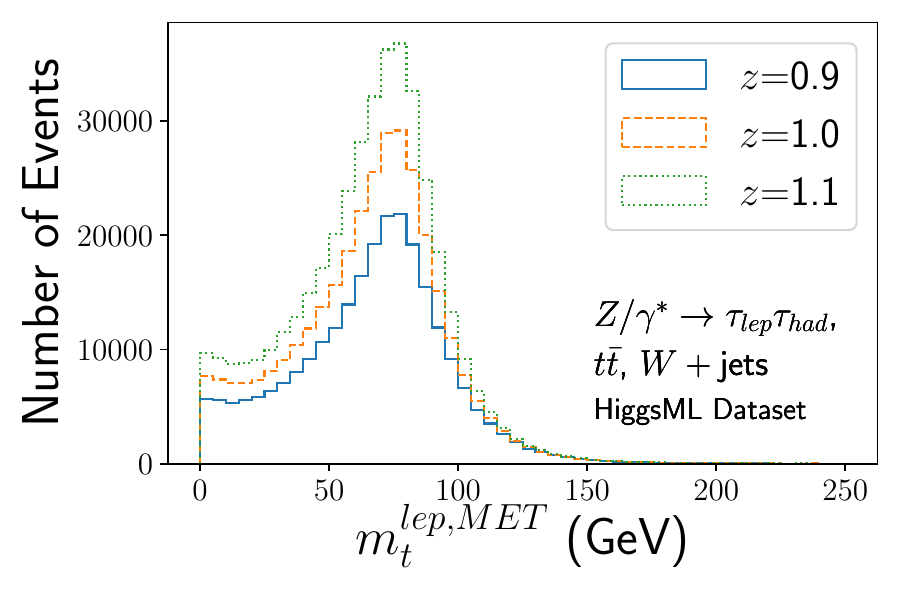}
    \caption{$m_t^{lep, MET} \text{ (GeV)}$}
    \label{fig:HiggsMLVariables:ptLepMET_0}
  \end{subfigure}
  \caption{Distribution of physics variables for three values of the nuisance parameter which controls the absolute tau lepton energy scale, $z=\{0.8,1,1.1\}$ for background processes. \subref{fig:HiggsMLVariables:TauPt_0} the transverse momentum of the hadronic $\tau$, \subref{fig:HiggsMLVariables:phiCentralityMET_0} the centrality in $\phi$ of the missing transverse energy vector with respect to the hadronic $\tau$ and the lepton, \subref{fig:HiggsMLVariables:ptLepMET_0} transverse mass of the missing transverse energy and the lepton. }
  \label{fig:HiggsMLVariables}
\end{figure}

At the nominal value of the nuisance parameter, $z=1$, the $\tau$~lepton energies are left unchanged. The impact of $z=0.9$ or $1.1$, on several features is shown in Fig.~\ref{fig:HiggsMLVariables}. The (unweighted) total number of events that pass the $E_\textrm{T}$ threshold for the $z=0.9$, $z=1$ and $z=1.1$ datasets are 618906, 719349 and 818201 respectively. The data are split into training and test set in the ratio 2:1. Since the data at various values of $z$ are generated from the nominal sample, the samples are to a large extent correlated. The train-test split therefore is determined before the skewing function and $E_\textrm{T}$ threshold are applied, ensuring complete independence between training and test sets.

Thirty bins are used to construct the template and observed histograms.

\begin{figure}[b!p]
  \centering
  \begin{subfigure}[b]{0.4\textwidth}
    \centering
    \includegraphics[width=\textwidth]{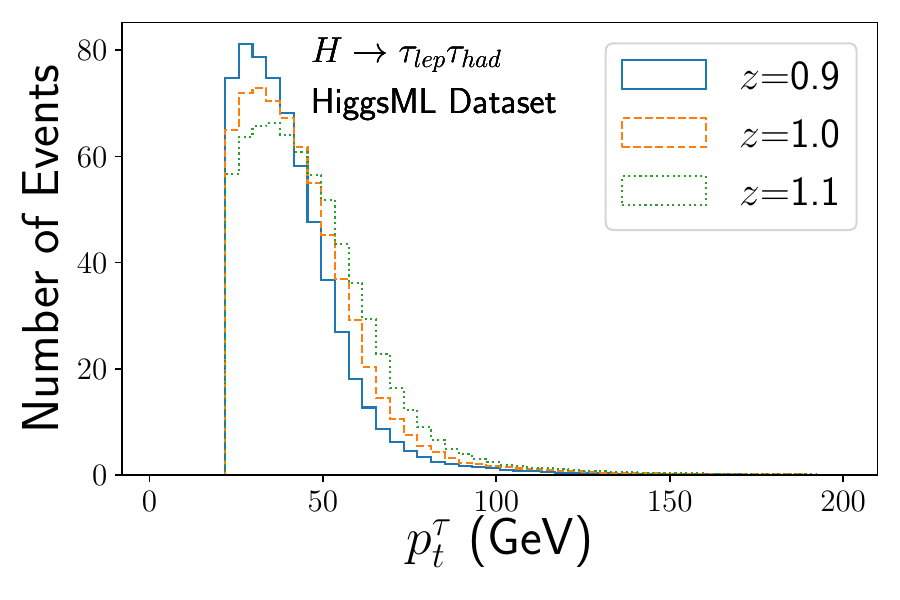}
    \caption{$p_t^\tau\text{ (GeV)}$}
    \label{fig:HiggsMLVariables:TauPt_1}
  \end{subfigure}
  \hfill
  \begin{subfigure}[b]{0.4\textwidth}
    \centering
\includegraphics[width=\textwidth]{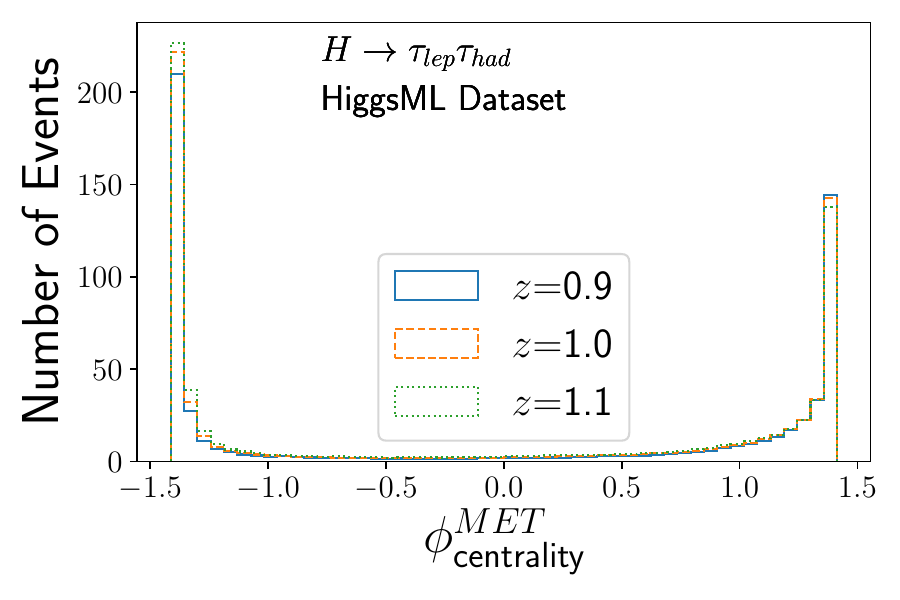}
    \caption{$\phi_{\text{centrality}}^{MET}$}
    \label{fig:HiggsMLVariables:phiCentralityMET_1}
  \end{subfigure}
  \hfill
  \begin{subfigure}[b]{0.4\textwidth}
    \centering
\includegraphics[width=\textwidth]{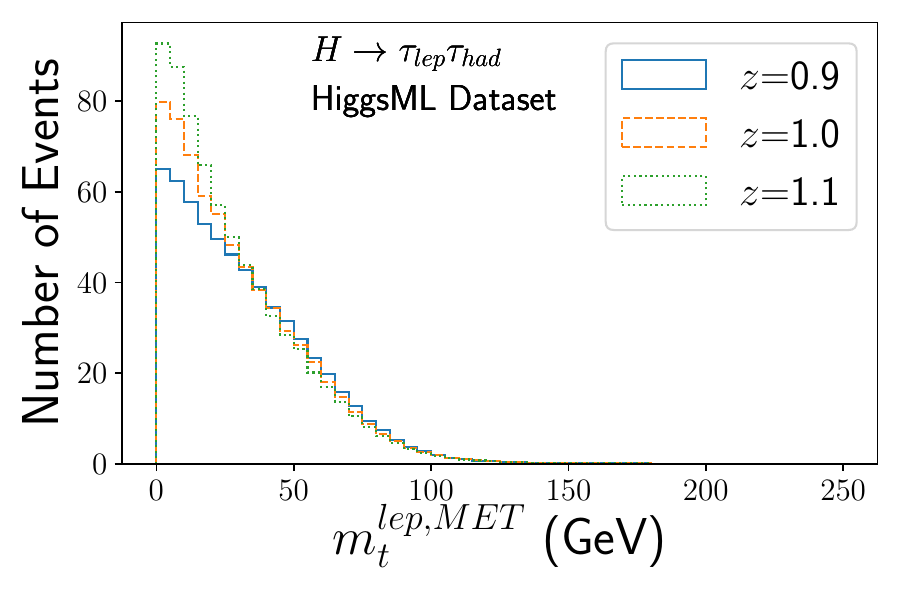}
    \caption{$m_t^{lep, MET} \text{ (GeV)}$}
    \label{fig:HiggsMLVariables:ptLepMET_1}
  \end{subfigure}
  \caption{Distribution of physics variables for three values of the nuisance parameter which controls the absolute tau lepton energy scale, $z=\{0.8,1,1.1\}$ for signal. \subref{fig:HiggsMLVariables:TauPt_1} the transverse momentum of the hadronic $\tau$, \subref{fig:HiggsMLVariables:phiCentralityMET_1} the centrality in $\phi$ of the missing transverse energy vector with respect to the hadronic $\tau$ and the lepton, \subref{fig:HiggsMLVariables:ptLepMET_1} transverse mass of the missing transverse energy and the lepton.}
  \label{fig:HiggsMLVariables_signal}
\end{figure}

\subsection{Description of Trained Models}
All methods were implemented using neural networks. The baseline classifier was trained only on data at $z=1$, while the data augmentation classifier, uncertainty-aware classifier and the adversarial classifier are all trained at 24 values spaced between $z=0.7$ and $z=1.4$. Two additional classifiers were also trained on data at $z=0.8$ and $z=1.1$ to estimate the best possible performance for an unparameterized classifier at these values of the nuisance parameter. 

Technical details about the training procedure and architectures of the models are given below.

\subsubsection{Baseline Classifier}
The neural network comprises 10 hidden layers with 512 nodes each, ReLU activations and L2 kernel regularizers for all but the first hidden layer and a final layer with a single node and sigmoid activation. It was trained with an RMSProp optimizer, BCE loss and a batch size of 4096.
\subsubsection{Data Augmentation}
The network comprises 10 hidden layers, each with 64 nodes, a ReLU activation, and L2 kernel regularizers for all but the first hidden layer and a final layer with sigmoid activation. The network was trained with an Adam optimizer~\cite{kingma2017adam}, BCE loss and a batch size of 4096.
\subsubsection{Adversarial Training}
Both the classifier and the adversary consist of 10 hidden layers with 64 nodes and ReLU activations and L2 kernel regularizers for all but the first hidden layer and a final layer with a single node with a sigmoid and linear activation for the classifier and adversary respectively. BCE is used as the classification loss and MSE as the adversarial loss. The two networks are attached with a gradient reversal layer which scales the gradient by $-0.2$ and trained with an RMSProp optimizer with $\lambda=1$ and a batch size of 4096.
\subsubsection{Uncertainty-Aware Classifier}
The uncertainty-aware classifier is comprised of two sub-networks combined with a custom `if-else' layer which outputs the result of the first sub-network if the input $z$ is less than $1$ and the result of the second sub-network otherwise. This  approach allows training of one sub-network longer than another if the performance in the $z\geq 1$  and $z<1$ regions converge at different rates.

Each sub-network consists of 10 hidden layers with 64 nodes, ReLU activations and L2 kernel regularizers for all but the first hidden layer and a final layer with a single node and sigmoid activation. They were trained using RMSProp optimizer, BCE loss and a batch size of 4096.

\subsubsection{Classifiers Trained On Data From True $z$}
Two neural networks were trained on data generated at $z={0.8}$ and $z={1.1}$ respectively. The network trained on data at $z={0.8}$ comprises 10 hidden layers with 64 nodes each, ReLU activations and L2 kernel regularizers for all but the first hidden layer and a final layer with a single node with a sigmoid activation. The network trained on data at $z={1.1}$ has an identical structure except that the hidden layers are 512 nodes wide. Both networks were trained using an RMSProp optimizer, BCE loss and a batch size of 4096.

\subsection{Results}
\begin{figure*}[b!p]
  \centering
  \begin{subfigure}[b]{0.3\textwidth}
    \centering
\includegraphics[width=\textwidth]{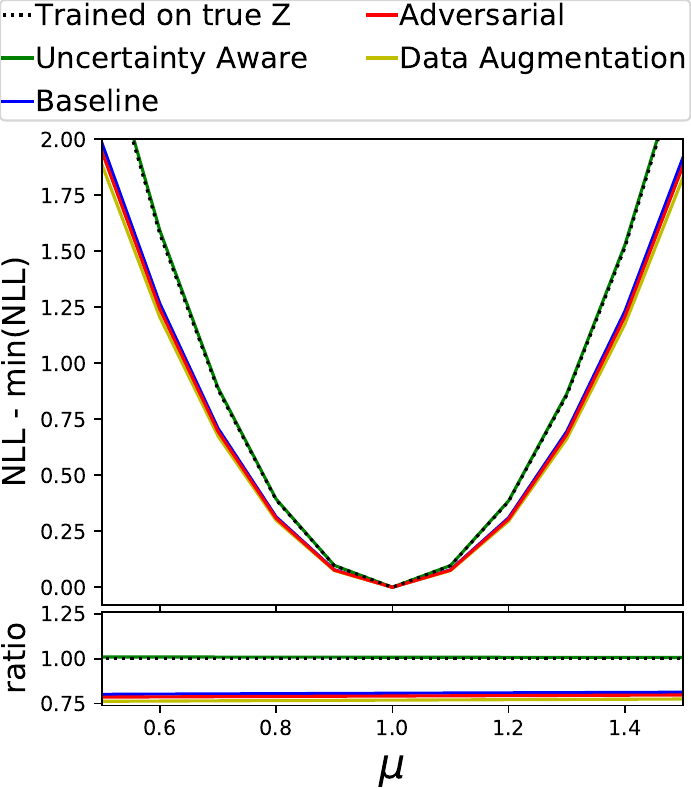}
    \caption{Systematic Down Data}
    \label{fig:1DNLLHiggsML:SystDown}
  \end{subfigure}
  \begin{subfigure}[b]{0.3\textwidth}
    \centering
    \includegraphics[width=\textwidth]{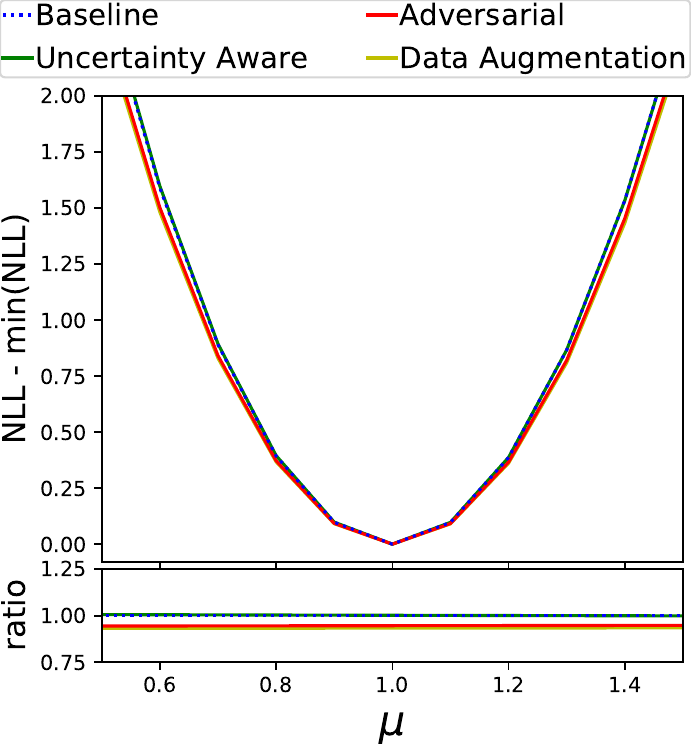}
    \caption{Nominal Data}
    \label{fig:1DNLLHiggsML:Nom}
  \end{subfigure}
  \begin{subfigure}[b]{0.3\textwidth}
    \centering
\includegraphics[width=\textwidth]{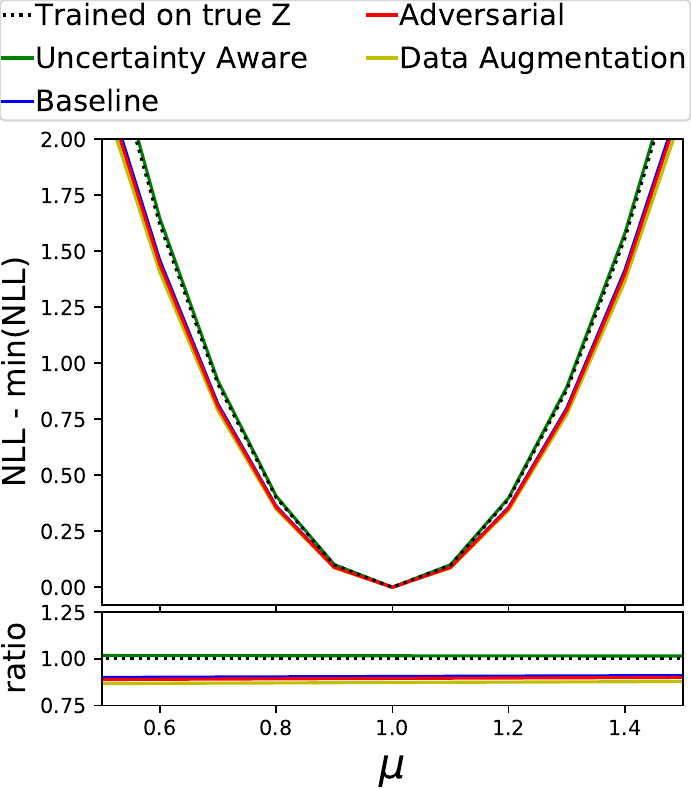}
    \caption{Systematic Up Data}
    \label{fig:1DNLLHiggsML:SystUp}
  \end{subfigure}
  \caption{Physics Dataset: Profiled NLL curves for all four classifiers evaluated on \subref{fig:1DNLLHiggsML:SystDown} systematic down ($z_\textrm{T}=0.8$), \subref{fig:1DNLLHiggsML:Nom} nominal data ($z_\textrm{T}=1.0$) and \subref{fig:1DNLLHiggsML:SystUp} systematic up data ($z_\textrm{T}=1.1$) where the true value of $\mu$ is 1. Narrower curves indicate more precise measurements having accounted for systematic and statistical uncertainties.}
  \label{fig:1DNLLHiggsML}
\end{figure*}

We evaluate the power of each method by examining the width of the profile likelihood curves in the parameter of interest, $\mu$, see Fig.~\ref{fig:1DNLLHiggsML}. The true value of $\mu$ was set to 1 (for comparisons when the true value of $\mu$ is set to 2 instead, refer to Appendix~\ref{sec:app:Physics:mu2}). When the data are generated with $z=1$, matching the assumption of the baseline classifier, we see that both the baseline and uncertainty-aware classifiers achieve the ideal performance, while the adversarial and data-augmented classifiers are slightly weaker. However, when the data are generated with a shift in the nuisance parameter ($z=0.8,1.1$) relative to the value used to build the baseline classifier, the uncertainty aware classifier maintains its ideal performance while the baseline classifier becomes less powerful.

\section{Conclusions}
\label{sec:conclusions}

In this paper\footnote{The code for this paper can be found at \url{https://github.com/hep-lbdl/systaware}}, we have advocated for uncertainty-aware classifiers where the dependence on nuisance parameter is maximized during training by exploiting parameterized classifiers~\cite{Cranmer:2015bka,Baldi:2016fzo}.  Using a Gaussian example and a realistic $H\rightarrow \tau\tau$ example, we have shown that the uncertainty-aware approach outperforms alternative methods that either are unaware of uncertainties or try to reduce the dependence on them during training.  Our approach is successful because it provides the most effective classifier for all values of the nuisance parameter.  This is useful when uncertainties are evaluated and when the nuisance parameter is profiled. It should be straightforward to apply this approach to multiple nuisance parameters although it was demonstrated on a single nuisance parameter in this paper.

While we advocate for maximally depending on the nuisance parameters, there could be cases where reducing the dependence could be beneficial.  For example, eliminating the dependence on a particular nuisance parameter reduces the analysis complexity. If the classifier is used to make a simple selection (cut-and-count), then reducing the dependence could also improve analysis sensitivity when the uncertainty on the nuisance parameters is large~\cite{Nachman:2019dol}.  For example, consider the case of an analysis that is considering adding a cut on a feature that is independent from all other features and that the cut has a signal efficiency $\epsilon_S$, a background efficiency $\epsilon_B$, and a relative uncertainty on $\epsilon_B$ that is $\delta$.  Clearly, one needs $\epsilon_S/\sqrt{\epsilon_B}\gtrsim 1$ for this cut to be useful. If $\delta$ is small, then the cut will certainly be useful.  If $\delta$ is large, it may be beneficial to not make the cut on the feature, which is the analog of rendering the analysis insensitive to $\delta$.  Caution must be taken in such cases because reducing the sensitivity to a nuisance parameter may hide the size of the true uncertainty.  This is the case for many two-point uncertainties such as hadronization modeling. Ultimately, the decision to use the additional feature or not depends on how the test statistic will be used in the analysis.  

A related topic to uncertainty-awareness is inference-awareness, where classifiers are trained using the final analysis objective as the loss.  The ultimate sensitivity will be achieved when both of these approaches are combined, which will be the topic of future investigation.

Uncertainty awareness is a relatively  straightforward extension of existing analyses performed at the LHC.  Many nuisance parameters can be varied without significant computational overhead and with the minimal changes we propose, analyses may be able to improve their sensitivity.  The biggest improvements are expected for analyses that are limited by experimental systematic uncertainties.  A growing number of analyses will fall into this category with the high-statistics of the High-Luminosity LHC and other HEP experiments across frontiers.


\section*{Acknowledgements}
AG thanks David Rousseau and Victor Estrade for fruitful discussions over several years, and for sharing code to generate the physics dataset, and also thanks Glen Cowan for fruitful discussions. We thank Tommaso Dorigo, Victor Estrade, David Rousseau and Kyle Cranmer for providing helpful comments on the manuscript.

BN is supported by the U.S. Department of Energy (DOE), Office of Science under contract DE-AC02-05CH11231. AG and DW are supported by the U.S.\ Department of Energy (DOE), Office of Science under Grant No. DE-SC0009920.

\input{main.bbl}
\appendix
\section{Likelihood Scans}
\label{sec:app:LikelihoodScans}
\begin{figure}[htp]
  \centering
  \begin{subfigure}[b]{0.23\textwidth}
    \centering
    \includegraphics[width=\textwidth]{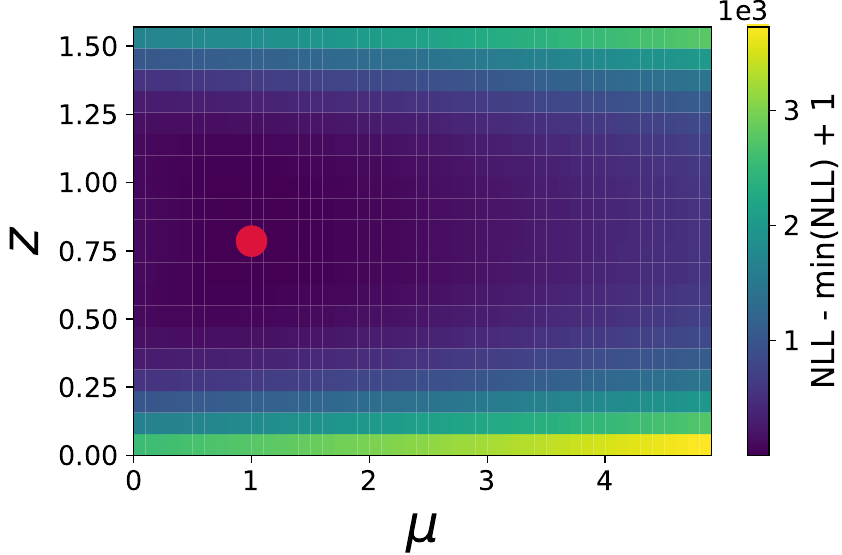}
    \caption{Data augmentation on data where $z=\frac{\pi}{4}$ }
  \end{subfigure}
  \hfill
  \begin{subfigure}[b]{0.23\textwidth}
    \centering
\includegraphics[width=\textwidth]{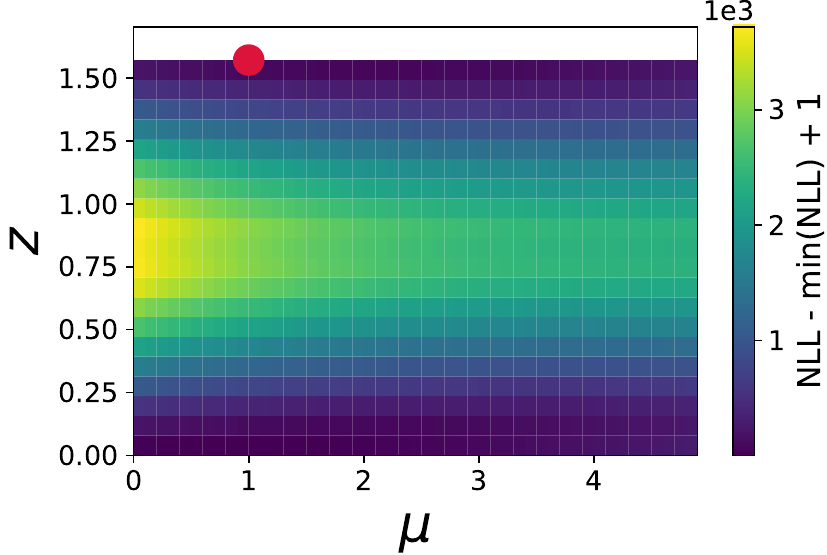}
    \caption{Data augmentation on data where $z=\frac{\pi}{2}$}
  \end{subfigure}
  \hfill
  \begin{subfigure}[b]{0.23\textwidth}
    \centering
\includegraphics[width=\textwidth]{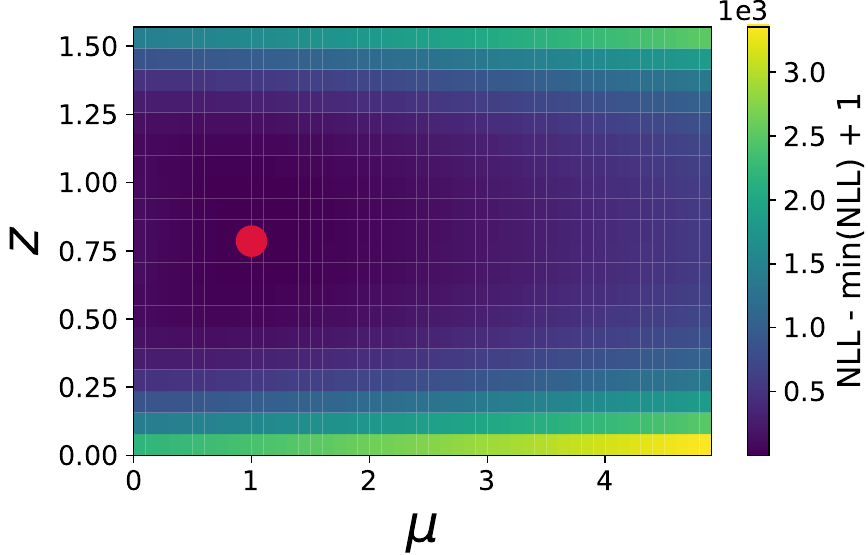}
    \caption{Adversarial training on data where $z=\frac{\pi}{4}$}
  \end{subfigure}
    \hfill
  \begin{subfigure}[b]{0.23\textwidth}
    \centering
\includegraphics[width=\textwidth]{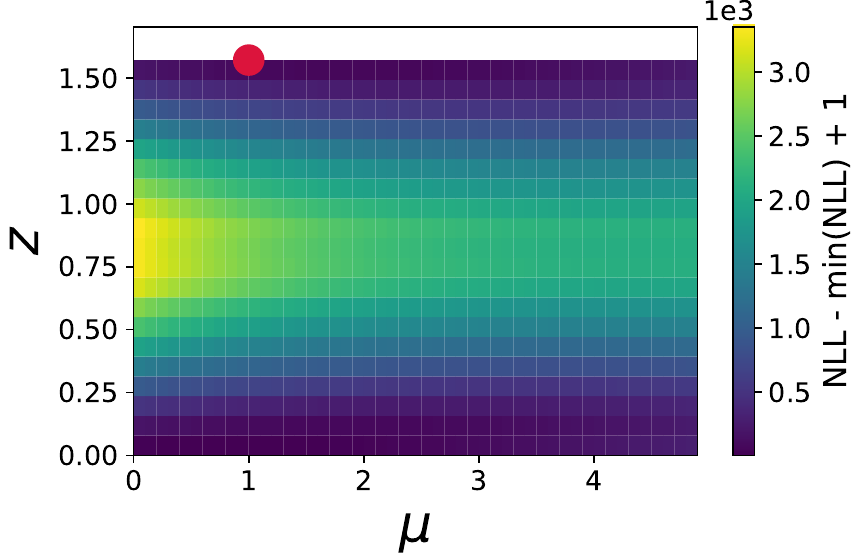}
    \caption{Adversarial training on data where $z=\frac{\pi}{2}$}
  \end{subfigure}
  \caption{The negative log-likelihood (Eq.~\ref{eq:2D_NLL}) as a function of the parameter of interest $\mu$ and the nuisance parameter $z$ for two example datasets in the Gaussian example, using templates from the data augmentation (top) and adversarial training (bottom).  On the left column, the data are generated with $z=\frac{\pi}{4}$, while in the right column, the data are generated with $z=\frac{\pi}{2}$. The red dot indicates the maximum likelihood estimate which coincides with the true value of $\mu, z$ in each case.}
  \label{fig:app:TwoDNLL:Toy}
\end{figure}

\begin{figure}[htp]
  \centering
  \hfill  
  \begin{subfigure}[b]{0.23\textwidth}
    \centering
    \includegraphics[width=\textwidth]{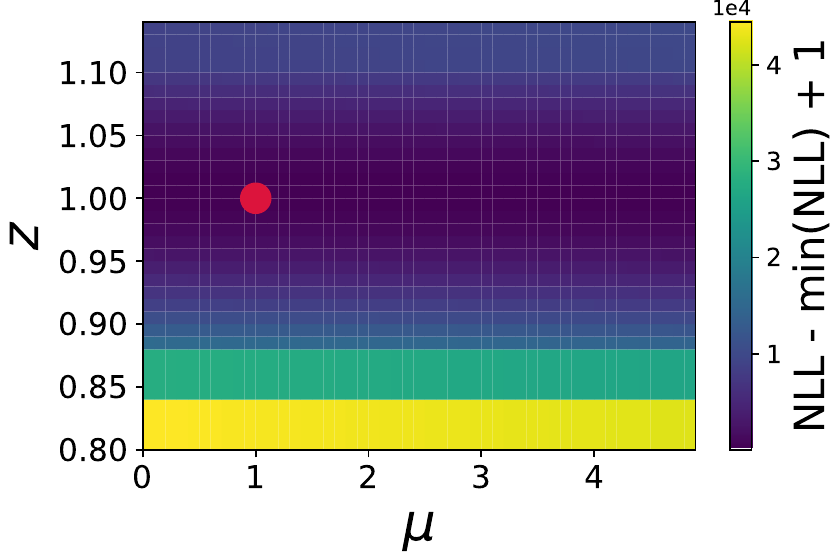}
    \caption{Baseline, on data where $z=1$ }
  \end{subfigure}
  \hfill
  \begin{subfigure}[b]{0.23\textwidth}
    \centering
\includegraphics[width=\textwidth]{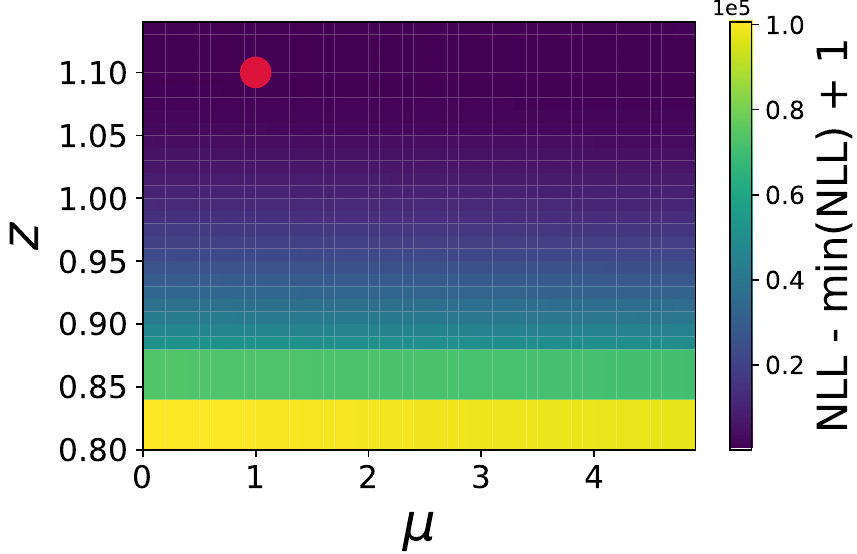}
    \caption{Baseline, on data where $z=1.1$}
  \end{subfigure}
  \hfill
  \begin{subfigure}[b]{0.23\textwidth}
    \centering
\includegraphics[width=\textwidth]{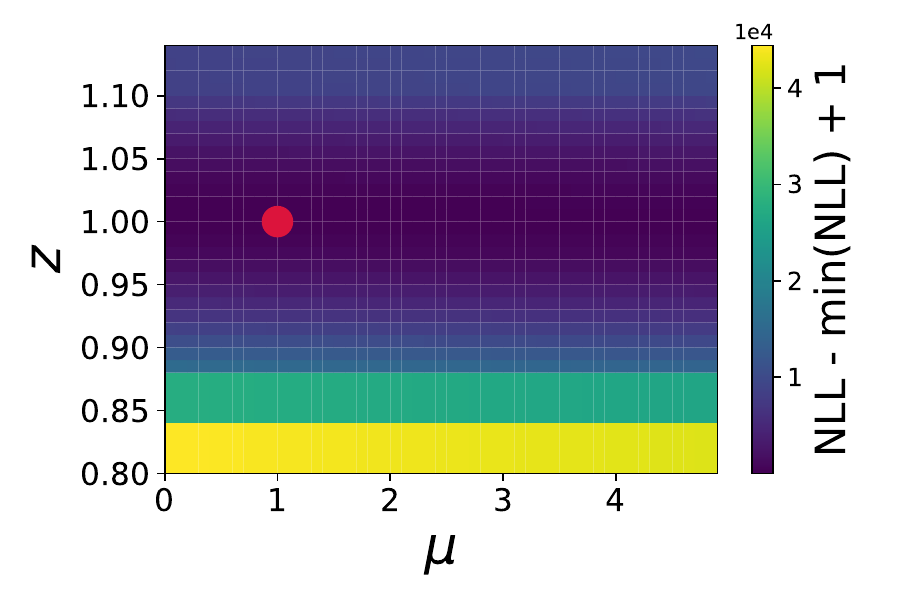}
    \caption{Uncertainty-aware, on data where $z=1$}
  \end{subfigure}
    \hfill
  \begin{subfigure}[b]{0.23\textwidth}
    \centering
\includegraphics[width=\textwidth]{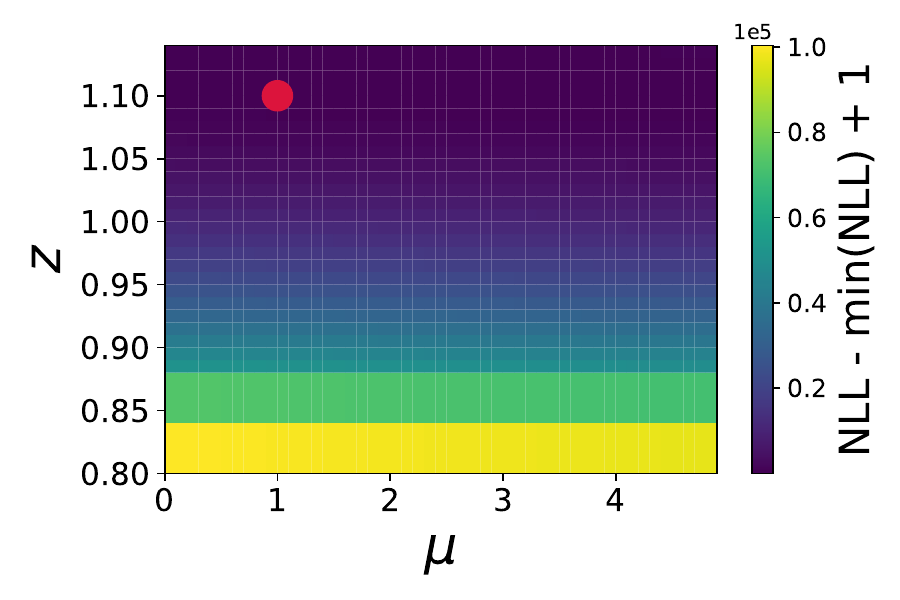}
    \caption{Uncertainty-aware, on data where $z=1.1$}
  \end{subfigure}
  \begin{subfigure}[b]{0.23\textwidth}
    \centering
    \includegraphics[width=\textwidth]{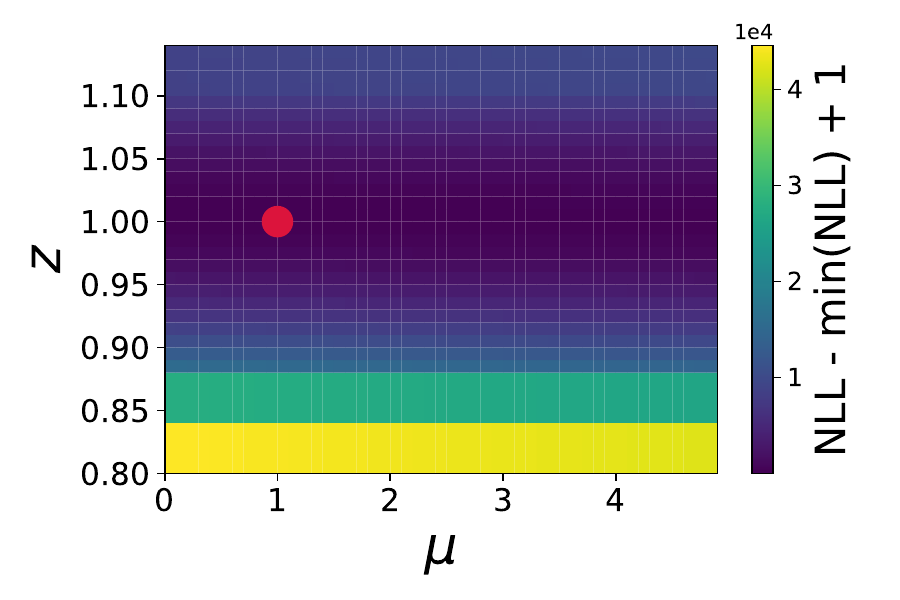}
    \caption{Data augmentation, on data where $z=1$ }
  \end{subfigure}
  \hfill
  \begin{subfigure}[b]{0.23\textwidth}
    \centering
\includegraphics[width=\textwidth]{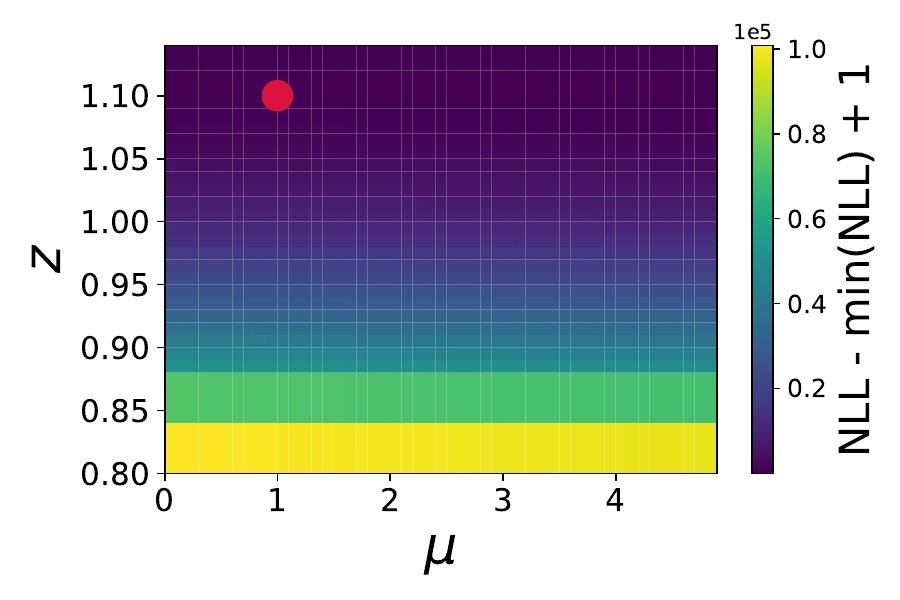}
    \caption{Data augmentation, on data where $z=1.1$}
  \end{subfigure}
  \hfill
  \begin{subfigure}[b]{0.23\textwidth}
    \centering
\includegraphics[width=\textwidth]{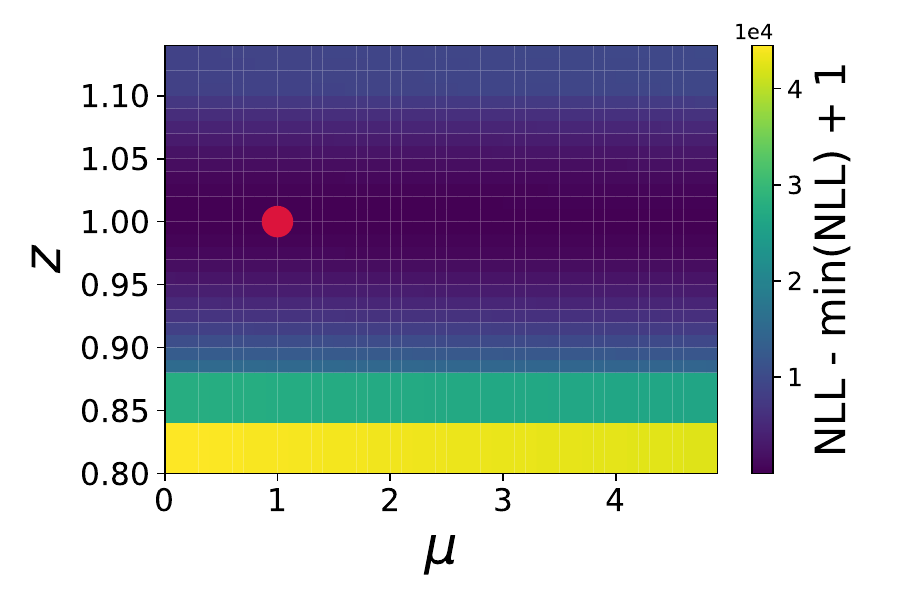}
    \caption{Adversarial training, on data where $z=1$}
  \end{subfigure}
    \hfill
  \begin{subfigure}[b]{0.23\textwidth}
    \centering
\includegraphics[width=\textwidth]{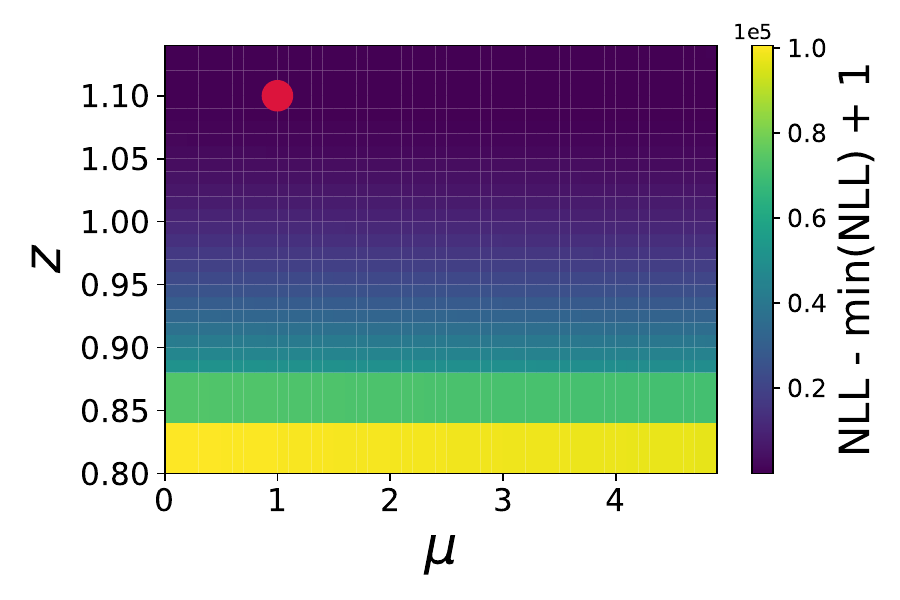}
    \caption{Adversarial training, on data where $z=1.1$}
  \end{subfigure}
  \caption{The negative log-likelihood (Eq.~\ref{eq:2D_NLL}) as a function of the parameter of interest $\mu$ and the nuisance parameter $z$ for two example datasets in the realistic example, using templates from the baseline (first row), systematic aware (second row), data augmentation (third row) and adversarial classifier (fourth row).  On the left column, the data are generated with $z=1$, while on the right column, the data are generated with $z=1.1$. The red dot indicates the maximum likelihood estimate which coincides with the true value of $\mu, z$ in each case. Note that the z-axis scale is not uniform in all figures.}
  \label{fig:app:TwoDNLL:HiggsML}
\end{figure}

Examples of the negative log-likelihood as a function of the parameter of interest $\mu$ and the nuisance parameter $z$ are shown for data augmentation and adversarial training in Fig.~\ref{fig:app:TwoDNLL:Toy} for the Gaussian example (Sec.~\ref{sec:Gaussian}). The same is shown for all four approaches compared for the realistic example (Sec.~\ref{sec:Physics}) in Fig.~\ref{fig:app:TwoDNLL:HiggsML}.

\FloatBarrier
\section{Gaussian Example with Auxiliary Measurement of $z$}
\label{sec:app:AuxMeasure}
A study is performed by replacing the prior on $z$ in Eq.~\ref{eq:2D_NLL} with a simultaneous auxiliary measurement. For simplicity the auxiliary measurement is of a Gaussian distribution with mean at $z_\textrm{T}$ and standard deviation of $0.5$. $10^5$ events uniformly weighted $0.1$ are generated at each of the 21 values of $z$. The negative log-likelihood then reads,
\begin{align}\nonumber
    -&\log{\mathcal{L}(\mu,z | \{x_i\})}\\ \label{eq:2D_NLL_aux}\nonumber &=-\sum^{n_\text{bins}}_{j=1}\bigg[ N_j\cdot \log{(\mu s_j + b_j)} - \mu s_j -b_j - \log(\Gamma(N_i)) \bigg] \\
    &\qquad-\sum^{m_\text{bins}}_{k=1}\bigg[ N^{aux}_k\cdot \log{(a^{z}_\text{k}) -a^{z}_\text{k} - \log(\Gamma(N^{aux}_k)} \bigg],
\end{align}
where $a^{z}_\text{k}$ is the number of events expected in bin $k$ of the auxiliary measurement for $z_\textrm{T}=z$ and $N^{aux}_k$ is the number of events actually observed in that bin. Four bins are used to construct the template and observed histograms for the auxiliary measurement.

\begin{figure}[htp]
  \centering
  \hfill  
  \begin{subfigure}[b]{0.23\textwidth}
    \centering
    \includegraphics[width=\textwidth]{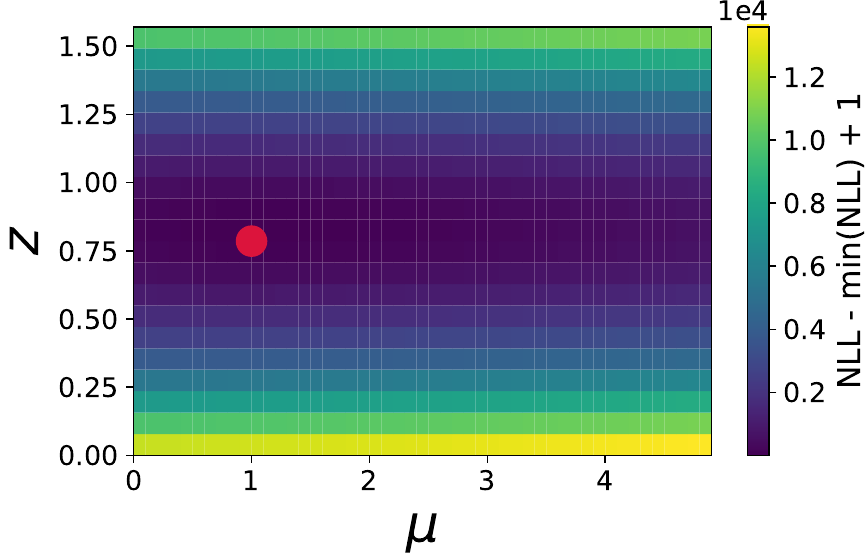}
    \caption{Baseline, on data where $z=\frac{\pi}{4}$ }
  \end{subfigure}
  \hfill
  \begin{subfigure}[b]{0.23\textwidth}
    \centering
\includegraphics[width=\textwidth]{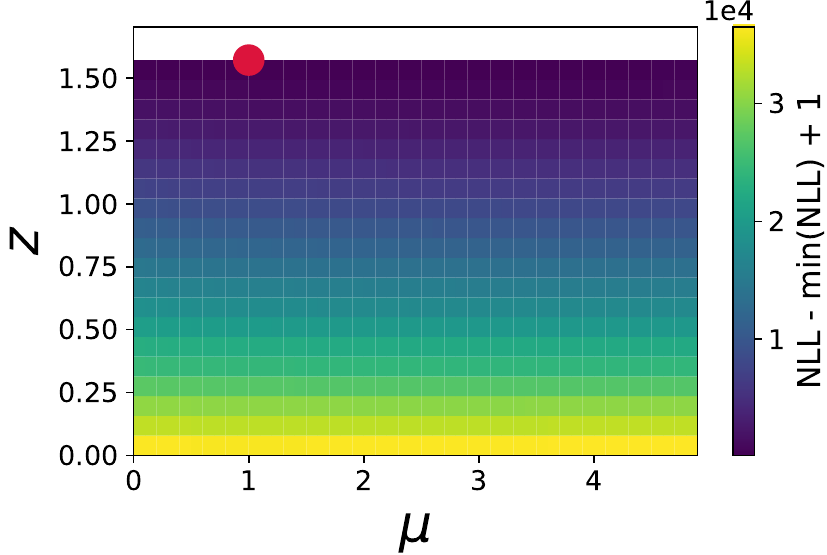}
    \caption{Baseline, on data where $z=\frac{\pi}{2}$}
  \end{subfigure}
  \hfill
  \begin{subfigure}[b]{0.23\textwidth}
    \centering
\includegraphics[width=\textwidth]{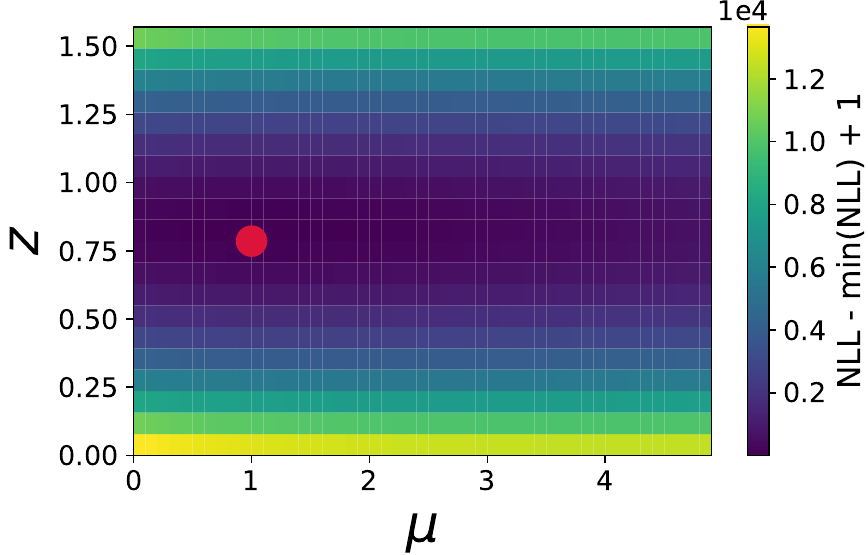}
    \caption{Uncertainty-aware, on data where $z=\frac{\pi}{4}$}
  \end{subfigure}
    \hfill
  \begin{subfigure}[b]{0.23\textwidth}
    \centering
\includegraphics[width=\textwidth]{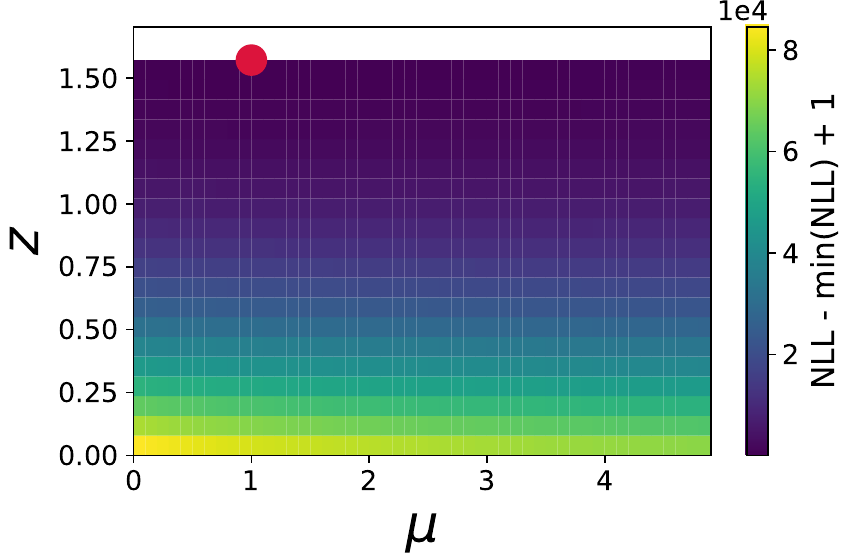}
    \caption{Uncertainty-aware, on data where $z=\frac{\pi}{2}$}
  \end{subfigure}
  \begin{subfigure}[b]{0.23\textwidth}
    \centering
    \includegraphics[width=\textwidth]{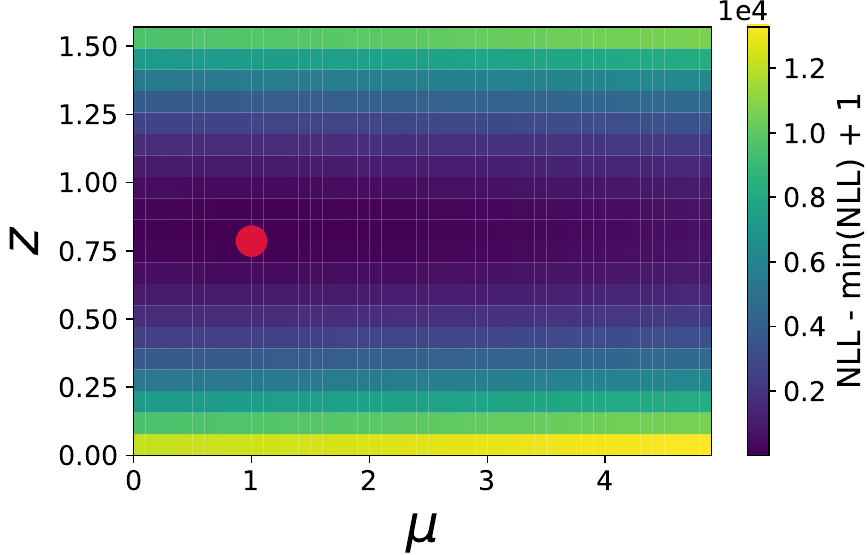}
    \caption{Data augmentation, on data where $z=\frac{\pi}{4}$ }
  \end{subfigure}
  \hfill
  \begin{subfigure}[b]{0.23\textwidth}
    \centering
\includegraphics[width=\textwidth]{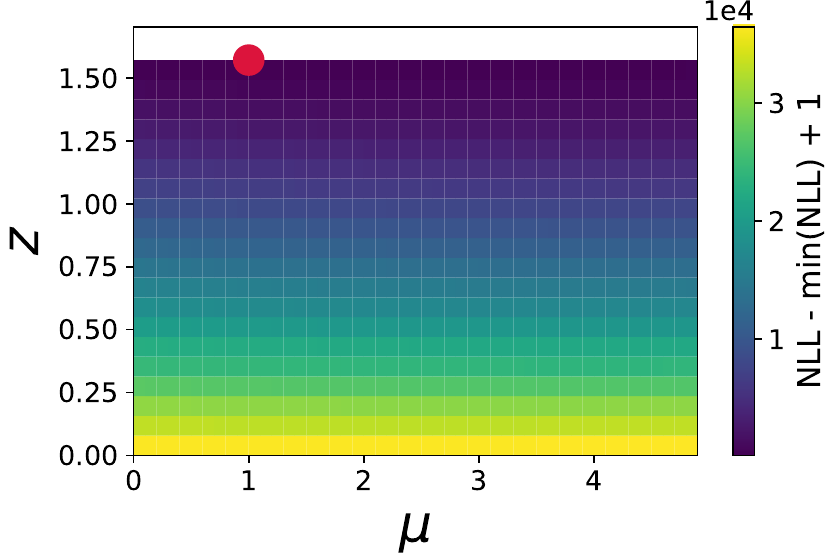}
    \caption{Data augmentation, on data where $z=\frac{\pi}{2}$}
  \end{subfigure}
  \hfill
  \begin{subfigure}[b]{0.23\textwidth}
    \centering
\includegraphics[width=\textwidth]{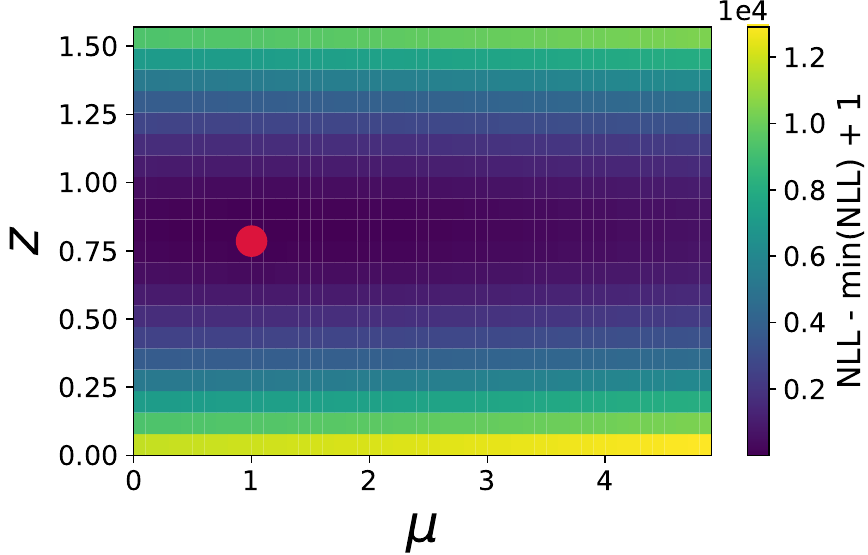}
    \caption{Adversarial training, on data where $z=\frac{\pi}{4}$}
  \end{subfigure}
    \hfill
  \begin{subfigure}[b]{0.23\textwidth}
    \centering
\includegraphics[width=\textwidth]{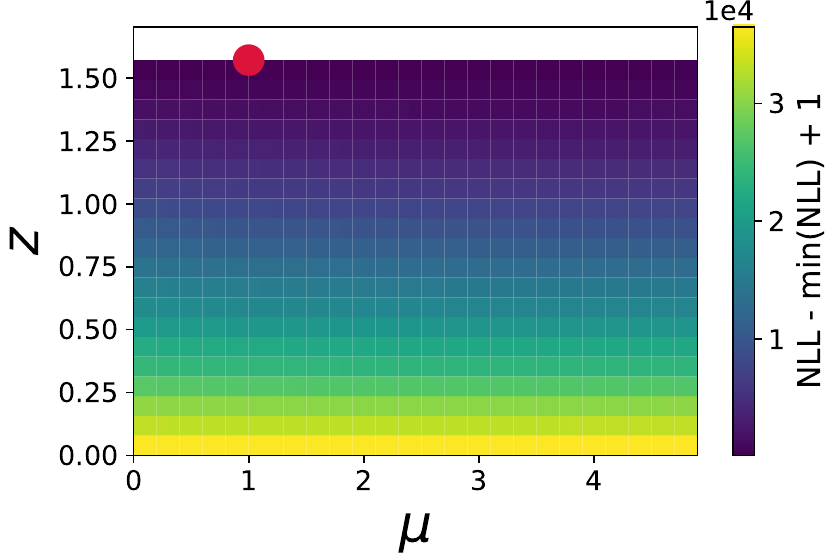}
    \caption{Adversarial training, on data where $z=\frac{\pi}{2}$}
  \end{subfigure}
  \caption{The negative log-likelihood (Eq.~\ref{eq:2D_NLL_aux}) as a function of the parameter of interest $\mu$ and the nuisance parameter $z$ in the auxiliary measurement study, using templates from the baseline (first row), systematic aware (second row), data augmentation (third row) and adversarial classifier (fourth row).  On the left column, the data are generated with $z=\frac{\pi}{4}$, while on the right column, the data are generated with $z=\frac{\pi}{2}$. The red dot indicates the maximum likelihood estimate which coincides with the true value of $\mu, z$ in each case. Note that the z-axis scale is not uniform in all figures.}
  \label{fig:2D_NLL_aux}
\end{figure}

\begin{figure}[htp]
  \centering
  \begin{subfigure}[b]{0.45\textwidth}
    \centering
    \includegraphics[width=\textwidth]{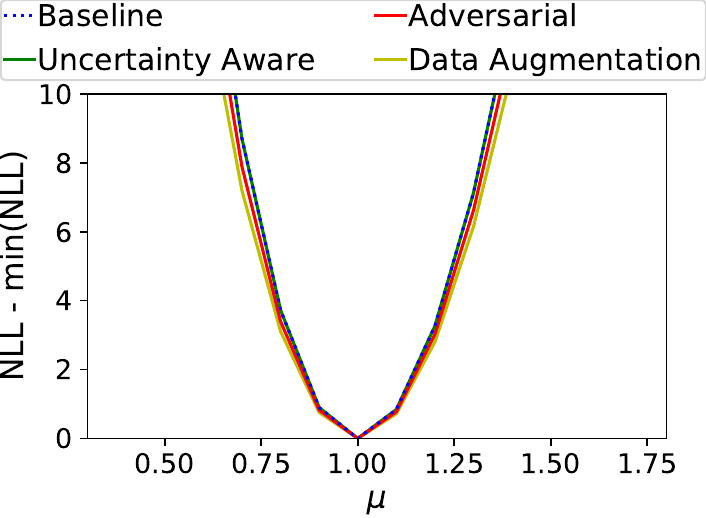}
    \caption{Data generated with $z=\frac{\pi}{4}$.}
  \end{subfigure}
  \hfill
  \begin{subfigure}[b]{0.45\textwidth}
    \centering
\includegraphics[width=\textwidth]{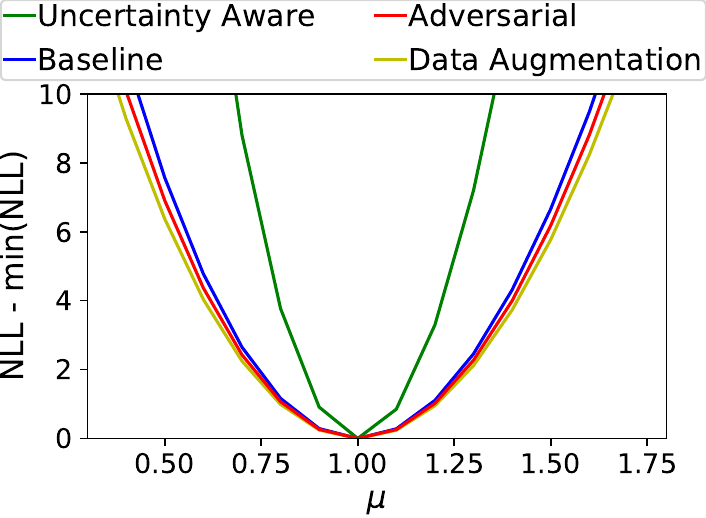}
    \caption{Data generated with $z=\frac{\pi}{2}$.}
  \end{subfigure}
  \caption{ The profile likelihood $\max_z\mathcal{L}(\mu,z)$ as a function of the parameter of interest, $\mu$ for likelihoods calculated with templates built from the various classifiers in the auxiliary measurement study. The baseline classifier assumes $z=\frac{\pi}{4}$, and matches the performance of the uncertainty-aware classifier in data generated with $z=\frac{\pi}{4}$ (top).  In data generated with $z=\frac{\pi}{2}$, the power of all classifiers other than the uncertainty-aware classifier become significantly weaker despite a better constraint on $z$ compared to Sec.~\ref{sec:Gaussian}.
    }
  \label{fig:profile_NLL_aux}
\end{figure}

The classifiers described in Sec.~\ref{sec:Gaussian} are re-used for this study, no re-training is required. The likelihood scans for the various approaches are shown in Fig.~\ref{fig:2D_NLL_aux}. For data generated at $z=\frac{\pi}{2}$ all approaches can exclude $z=0$ since the auxiliary measurement constrains $z$ much more than the prior used in Sec.~\ref{sec:Gaussian}; Fig.~\ref{fig:2DNLLToy}. The profile likelihood in Fig.~\ref{fig:profile_NLL_aux} shows that although the curves are narrower compared to Fig.~\ref{fig:1DNLLToy}, the overall conclusions discussed in Sec.~\ref{sec:Gaussian} remain valid.

\section{Tests at $\mu=2$ for Physics Example}
\label{sec:app:Physics:mu2}
\begin{figure*}[b!p]
  \centering
  \begin{subfigure}[b]{0.3\textwidth}
    \centering
\includegraphics[width=\textwidth]{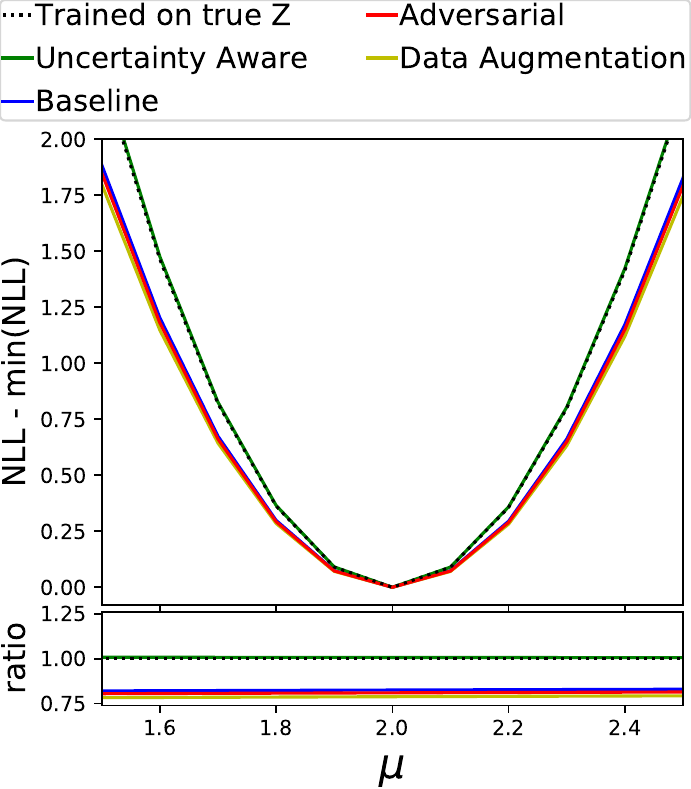}
    \caption{Systematic Down Data}
    \label{fig:1DNLLHiggsML:SystDown:mu2}
  \end{subfigure}
  \begin{subfigure}[b]{0.3\textwidth}
    \centering
    \includegraphics[width=\textwidth]{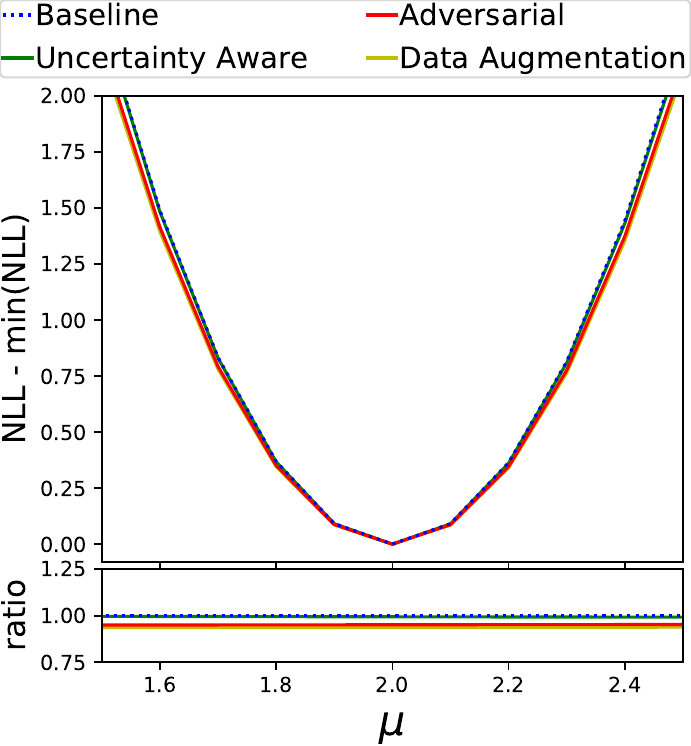}
    \caption{Nominal Data}
    \label{fig:1DNLLHiggsML:Nom:mu2}
  \end{subfigure}
  \begin{subfigure}[b]{0.3\textwidth}
    \centering
\includegraphics[width=\textwidth]{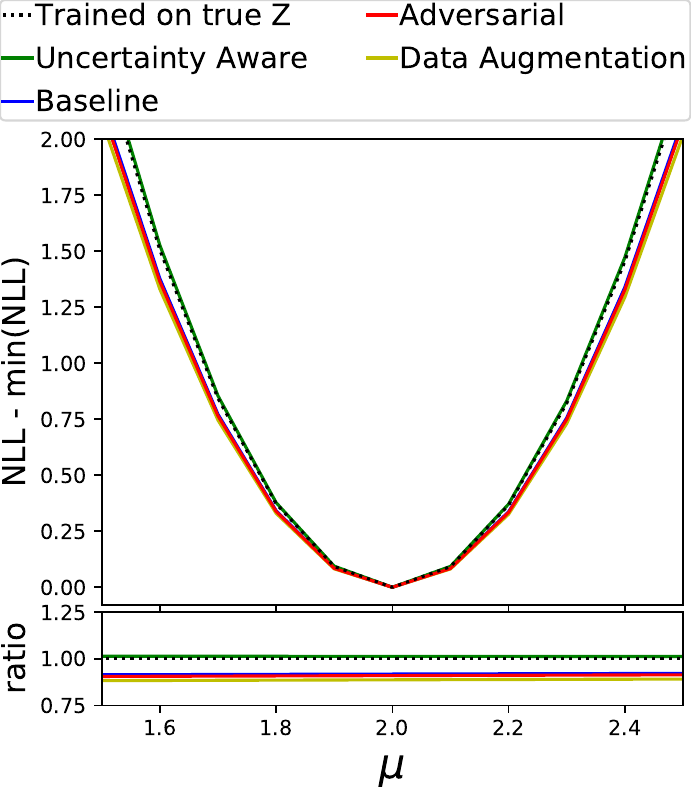}
    \caption{Systematic Up Data}
    \label{fig:1DNLLHiggsML:SystUp:mu2}
  \end{subfigure}
  \caption{Physics Dataset: Profiled NLL curves for all four classifiers evaluated on \subref{fig:1DNLLHiggsML:SystDown:mu2} systematic down ($z_\textrm{T}=0.8$), \subref{fig:1DNLLHiggsML:Nom:mu2} nominal data ($z_\textrm{T}=1.0$) and \subref{fig:1DNLLHiggsML:SystUp:mu2} systematic up data ($z_\textrm{T}=1.1$) where the true value of $\mu$ is 2. Narrower curves indicate more precise measurements having accounted for systematic and statistical uncertainties.}
  \label{fig:1DNLLHiggsML:mu2}
\end{figure*}

The comparison of the four approaches was also performed for data where the true value of the parameter of interest $\mu$ is 2. The profile likelihoods in Fig.~\ref{fig:1DNLLHiggsML:mu2} show that the conclusions of Sec.~\ref{sec:Physics} remain valid.

\end{document}

%% file: main.bbl
%